\documentclass[astrosymb]{aastex701}
\usepackage{subcaption}
\shorttitle{Hubble Tomography via Linear Cosmography}
\shortauthors{B. Bochner and A. Jin}

\begin{document}

\title{Exploring Cosmological Tensions with Hubble Parameter Tomography via Linear Cosmography}

\author[orcid=0000-0002-4512-7617,gname='Brett',sname='Bochner']{Brett Bochner}
\affiliation{Department of Physics and Astronomy, Hofstra University, NY, USA}
\email[show]{Brett.D.Bochner@hofstra.edu, brett\_bochner@alum.mit.edu}  

\author[gname='Aiden',sname='Jin']{Aiden Jin}
\affiliation{John L. Miller Great Neck North High School, NY, USA}
\email{aidenjin24@gmail.com}

\begin{abstract}
Given the persistence of various tensions in the ``Cosmic Concordance" -- such as the ``Hubble Tension", and possible departures from $\Lambda$CDM time evolution -- seen from combinations of complementary data sets (e.g., Cosmic Microwave Background, Baryon Acoustic Oscillations, Type Ia Supernovae), it remains an ongoing possibility for these to have a real cosmological origin. If one assumes such deviations to be real, a model-independent formalism (cosmography) is useful for locating the source of the problem with concordance cosmology. The extraordinarily good fit of $\Lambda$CDM to CMB data shows that it was a successful model of the universe at high redshift. Yet at lower redshift -- when the dark energy density becomes significant, and its precise physical nature becomes important -- the universe may have gone off the track of simple $\Lambda$CDM. Here we use linear cosmography fits to binned Supernova data to reconstruct the detailed temporal history of the Hubble parameter, probing for interesting time-dependent behaviors of the expansion rate during and after the onset of cosmic acceleration. Using combined Type Ia supernovae from the Dark Energy Survey 5-Year data release and the Union2.1 compilation, we find intriguing hints of oscillations in the Hubble parameter during the acceleration era. While these hints are of limited statistical significance, and not robust under different redshift binnings, we present this work as a proof-of-concept demonstration of this method for reconstructing the Hubble parameter evolution, which may be useful for the voluminous Supernova data sets anticipated to become available during the next few years.
\end{abstract}



\section{Introduction and Motivation \label{SecIntro}} 

Following the discovery of the cosmic acceleration about a quarter of a century ago \citep{RiessAccel98,PerlAccel99}, once the fact of a negative deceleration parameter was sufficiently well established, attention zeroed in on determining the parameters of the most obvious model of an accelerating universe: $\Lambda$CDM. This straightforward extension of the standard matter-dominated cosmology re-introduced Einstein's ``Cosmological Constant", $\Lambda$, to the mix of baryonic and (probably Cold) Dark Matter, in the right amount needed to achieve a spatially flat universe, as indicated by Cosmic Microwave Background Radiation (CMB) observations \citep{BoomFlat}. For much of the following decades, as the age of ``Precision Cosmology" arrived\footnote{Note, though, that the terms ``Precision Cosmology" \citep{TurnerPrecCos} and ``Cosmic Concordance" \citep{OstrikerConcord} were actually used as early as the 1990's.} with powerful results from multiple data sets -- providing constraints on complementary combinations of parameters -- it appeared that we were narrowing down into a ``Cosmic Concordance" : a tightly-constrained region of $\Lambda$CDM parameter space agreeing with all major observations, on the way towards confirming the Cosmological Constant paradigm \citep[e.g., Figures 5-7 of][]{SCPUnion21}.

As with all parameter space searches, where statistics are limited and significant gaps and systematics exist in the data, anomalies and tensions eventually emerged in a variety of ways, popping up as apparent inconsistencies with the $\Lambda$CDM Concordance Model \citep[various problems and suggested solutions explored exhaustively in][]{ChallengesLCDM,CosmoVerse}. For each such issue, the debate over whether it is resolving itself naturally with further observations, or persisting, or even worsening, can be contentious. But when it becomes clear that inconsistencies are not disappearing with additional data, then only two possibilities remain: either pernicious systematics are corrupting the conclusions; or the basic theoretical paradigm at the core of the model is incomplete (or even incorrect). Since the main focus of this paper is the evaluation of the temporal history of the expansion rate of the universe over time, we will be mainly concerned with two well-known tensions: the ``Hubble Tension" \citep[e.g.,][]{HubbTension}; and the possibility of active departures from $\Lambda$CDM evolution, as indicated by Baryon Acoustic Oscillation (BAO) observations from the Dark Energy Spectroscopic Instrument Data Release 2 (DESI DR2) \citep{DESIBAO}.

The much-discussed Hubble Tension, is the result that \textit{direct} measurements of the Hubble Constant --  i.e., the limit of Hubble Parameter $H(z)$ as $z \rightarrow 0$ -- performed at low-redshift using Type Ia Supernovae (SNIa) or other standard candles or rulers, do not match the values of the Hubble Constant \textit{indirectly inferred} from various higher-redshift measurements (most notably, of the CMB). The other major tension we refer to, a deviation from Concordance -- specifically, a preference for dynamical dark energy over $\Lambda$CDM evolution at the $2.8-4.2\sigma$ level (depending upon which SNIa data set was used in conjunction to produce constraints) -- was observed by DESI DR2, a re-confirmation (and strengthening) of their Data Release 1 results.

First of all, it must be noted that while these two tensions are often discussed in the cosmological community as being two separate problems -- likely due to the observational reality that the measurements are made using different types of astronomical objects, and done largely by different scientific collaborations -- they are nevertheless aspects of the \textit{same problem}, from a theoretical point of view. If we begin with the hypothesis that these tensions are \textit{real}, and not just a result of differing systematics from different types of observational measurements, then each problem generally implies the other one. (Additional discussion of the matter is given in Appendix~\ref{BothSameProb}.)

Much of the recent Hubble Tension debate has centered upon the question of systematics in the anchoring of the cosmic distance ladder for the Type Ia Supernovae (SNIa). This dispute has involved the question of using either Cepheid variable stars or other standard candles -- such as ``Tip of the Red Giant Branch" (TRGB) stars, and J-region Asymptotic Giant Branch star (JAGB) stars -- for calibrating the SNIa \citep{FreedmanCephSys}; and differing results have been obtained from competing groups even in cases where both are using Cepheids as the SNIa anchor. The SH0ES Team \citep{RiessH0} finds a value of $H_{0} = 73.30 \pm 1.04$~km~sec$^{-1}$ Mpc$^{-1}$, $5\sigma$ away from the Planck CMB results \citep{PlanckH0}. In contrast, the Chicago-Carnegie Hubble Program (CCHP) group \citep{FreedmanH0} claims consistency with the CMB-derived value of $H_{0}$; still, while the CCHP and Planck results may be within the error bars of each other, the estimated $H_{0}$ from CCHP is still higher than that from the CMB by 3~km~sec$^{-1}$ Mpc$^{-1}$ (70.4 versus 67.4~km~sec$^{-1}$ Mpc$^{-1}$). Thus, while SH0ES and CCHP may differ on the precise level of the SNIa tension with CMB-derived $H_{0}$ constraints, both collaborations still measure \textit{higher} values of $H_{0}$ than that from Planck. (And the latest DESI results only strengthen the case for tension with Concordance $\Lambda$CDM.)

Therefore, for the purposes of this paper, we will make the presumption that the observed deviations from Concordance are real, and ask two questions: (1) When in cosmic history are such deviations most likely to occur? (2) What method is best for pinning down the ``epoch of deviations", so to speak?

Regarding the first question, we must consider the role of dark energy at the time of the last scattering, and its effects upon CMB observations. If we disregard the ``Early Dark Energy" models (seemingly not favored, \citet{ACT6noEDE}), then at these very high redshifts, the dark energy is completely sub-dominant -- $\Omega^{CMB}_{\Lambda} \sim 10^{-9}$ ! -- and thus it is essentially physically impotent during that epoch. Its main role in CMB observations is to help determine the distance to the last scattering surface (which affects the observed angular size of the sound horizon at recombination), and also to translate the \textit{physical densities} actually measured by CMB observations (e.g., baryon density $\rho _{b}$), into indirectly-estimated values of theoretical model parameters: e.g., $\Omega _{b} = \rho _{b}/h^{2}$, where $h \equiv H_{0}$/(100~km~sec$^{-1}$ Mpc$^{-1})$. As detailed in Appendix~\ref{BothSameProb}, all of these quantities are specifically designed to have meaning the context of $\Lambda$CDM. But CMB-derived parameters involving the dark energy only take on physical meaning when one is patching together the entire cosmic history, including times when the dark energy density is cosmologically significant. Thus, the precise physical nature of the dark energy is almost entirely irrelevant during the CMB epoch, since it doesn't ``do anything" at those early cosmic times.\footnote{The only ``direct" impact of the physics of dark energy upon CMB observations involves the Late ISW effect, affecting our observations at low CMB multipoles; this is the least well-understood part of the CMB power spectrum in any case, beset by several anomalies -- including an anomalously strong ISW effect, as compared with expectations from $\Lambda$CDM \citep{ChallengesLCDM}.}

Despite the aforementioned tensions with Concordance cosmology, CMB observations have nevertheless been extremely consistent with the $\Lambda$CDM paradigm. These include the Atacama Cosmology Telescope Data Release 6 (ACT DR6) results \citep{ACTDR6}, for multipoles $l \gtrsim 500$ or so; and similarly for the results from Planck, and elsewhere. Thus $\Lambda$CDM seems to be a very good model -- or at least, a very good empirical ``stand-in" -- for the true cosmological evolution, at very high redshift. But, as just explained, the specific details of dark energy physics were not materially relevant at such early times; therefore, we would be unsurprised to see a model with Cosmological Constant dark energy fit the CMB data very well, even if $\Lambda$CDM is ultimately \textit{wrong}, as long as the phenomenological model being used to represent the dark energy has parameters (e.g., $\Omega _{\Lambda} \simeq 1 - \Omega_{M}$) that are appropriately well chosen.

The distance to the last scattering surface is given via $d_{\mathrm{CMB}} \propto H_{0}^{-1} \int a(t) ^{-1} dt$; any deviation from the $a(t)$ expected from $\Lambda$CDM could be ``swallowed up" by instead changing our CMB-estimated value of $H_{0}$, thus yielding an apparent Hubble Tension -- and perhaps this is exactly what we have seen. The real problem may therefore lie in the extrapolation of high-redshift CMB data down to low-redshift. Much of the discussion of the Hubble Tension has been rooted in a discussion of conflicting results from different observational techniques, beset by differing systematics \citep[e.g.,][]{FreedmanCephSys}; but perhaps the real culprit is a function of the \textit{cosmological epoch} of the observations, with a conflict between the high-redshift versus low-redshift universe. In that hypothesis, the flaws with the Cosmological Constant model would only appear as the explicit physical nature of the dark energy becomes important, once its density ceases to be negligible; in other words, around (or shortly before) the onset of cosmic acceleration, which occurs near $z \sim 0.63-0.66$ (depending upon which $\Lambda$CDM parameter estimates are used). Thus $\Lambda$CDM would appear perfectly fine at high redshift, but would begin to fail for $z \lesssim 0.7$ or so. As will be discussed below, it is already apparent with BAO data alone \citep{DESIBAO} that a strong ``jerk" in the expansion rate occurred at just around this epoch, with a suddenly-larger $H(z)$ than that expected from $\Lambda$CDM; and 
the inclusion of SNIa data will allow us to extend the analysis of the Hubble Parameter evolution down to lower redshifts, where the dark energy really begins to dominate, and the details of its physical nature become paramount.

Regarding the second of our above questions: much as a doctor seeks to diagnose a malady by probing the patient and asking, ``Where does it hurt?", one can imagine ``feeling along" the history of $H(z)$ in order to pinpoint exactly when deviations from Cosmological Constant dark energy (\textit{if} they exist) first created significant disturbances in the cosmic expansion rate. Barring clear-cut theoretical guidance in the case of non-Concordance, a model-independent approach like Cosmography would seem to be a useful first step. Many interesting cosmographic studies have been done to date, including by one of us \citep{BochnerLamDev}; but such studies often perform expansions about $z = 0$ (not the most cosmologically interesting pivot-point), they smear any interesting behaviors over the whole redshift range being modeled (instead of pinpointing them), and require so many cosmographic parameters (to model the whole redshift range) that the uncertainties on those parameters become huge. In essence, it is difficult to provide enough (precisely estimated) cosmographic parameters to accurately model an unknown but detailed theoretical curve; a problem we explored in \citet{BochnerLamDev}, and also studied elsewhere \citep[e.g.,][]{CosmogNoGood}. Therefore, instead of performing a general cosmographic expansion -- and to be more focused on precisely locating interesting temporal behaviors -- we will bin the SNIa data in redshift, and demonstrate how a fairly elementary procedure employing straight-line cosmographic fits to small (thus necessarily \textit{linearized}) sections of the data can be used to reconstruct a (binned version) of the Hubble Parameter over the entire redshift range covered by the SNIa data.

Evaluations of binned SNIa to estimate deviations from $\Lambda$CDM are not new, with the most obvious applications being redshift-binned estimates of the dark energy equation of state $w(z)$ \citep[e.g.,][]{SCPUnion21}, or of directly measuring deviations in the distance modulus $\mu (z)$ \citep[e.g.,][]{DES5YrSNIa,HubbTensTomog}. Nevertheless, we believe that providing a binned reconstruction of the Hubble Parameter is particularly illuminating, given that it directly probes the Hubble Tension by connecting $H(z)$\footnote{As explained in the next section, it is mathematically more sensible to work in $y$-redshift, instead of the traditional ($z$)-redshift; we therefore design our binnings in order to best reconstruct $H(y)$ -- rather than $H(z)$ -- although the two are formally equivalent.} between the high-redshift and low-redshift ends of the spectrum of data sources; and also, the Hubble Parameter provides an immediate visualization for how the universe ``speeds up" or ``slows down" in its acceleration, both in the absolute sense, and also relative to $\Lambda$CDM.

A parallel idea of using redshift binning to reconstruct the cosmic history of the Hubble (and other) tensions was explored in \citet{LocH0Tensions}, where they discussed the idea of looking for differences in the $z \rightarrow 0$ extrapolated value of the Hubble constant $H_{0}$, when obtained from data at different redshifts; these extrapolations would be performed using the $\Lambda$CDM model, and a conflict of $H_{0}$ could thus be interpreted as an inconsistency in that standard cosmological model. In our work here, however, we prefer to avoid assuming any model at all, and instead employ a method that is \textit{temporally local} at all times -- a method ``without a memory" of cosmological history, operating completely independently at every redshift.

In the ensuing sections, we will use our binned, linear cosmography approach to explore the time dependence of the cosmic expansion rate, and find intriguing results -- including hints of what appears to be an oscillatory pattern in the Hubble Parameter during the post-acceleration era of the universe. While such hints are still at relatively low levels of statistical significance -- and not yet robustly independent of one's choice of redshift binning -- we believe that the results presented here are an interesting ``proof-of-concept" demonstration of this method, which we hope will be used extensively for analyses of the vast samples of new SNIa data expected from surveys such as LSST \citep{SNeFromLSST}.

\section{Methodology \label{Methods}} 

Proceeding with the derivation of our method, we begin with the well known formula for Luminosity Distance $d_{L}$ \citep{KolbTurner}, assuming spatial flatness ($k = 0$): 
\begin{equation}
	d_{L} (t) = (1 + z) ~ a_{0} ~ r(t) = (1 + z) ~ a_{0} ~ \int^{t_{0}}_{t} \frac{c}{a(t^{\prime})} dt^{\prime} ~,
	\label{EqndLumVSt}
\end{equation}
where $a_{0} \equiv a(t_{0})$ is the present-day ($t \equiv t_{0}$) value of cosmic scale factor $a(t)$, and $r(t)$ is the coordinate distance to the SNIa emitting the signal at time $t$ (with measured redshift $z$). With the definition of the redshift, $z \equiv [a_{0}/a(t)] - 1$, we have: $dz/dt = - [a_{0}~\dot{a}(t)] / [a(t)^2] \equiv - [a_{0}/a(t)] ~ H(t)$. We may thus rewrite the differential: $dz = dt \cdot (dz/dt) = - [a_{0}/a(t)] \cdot H(t) \cdot dt$. Re-expressing the Hubble Parameter as a function of redshift, $H(t) \rightarrow H(z)$, we see that Eq.~\ref{EqndLumVSt} can be equivalently given as:
\begin{equation}
	d_{L} (z) = c ~ (1 + z) ~ \int^{z}_{0} \frac{1}{H(z')} dz' ~.
	\label{EqndLumVSz}
\end{equation}

\subsection{The Redshift Variables \label{yVSzRedshift}}

At this point, we must note that while redshift $z$ is the traditional variable most often used in cosmological (and cosmographic) analyses, it is no longer mathematically justifiable, given the modern availability of high-redshift ($z \gtrsim 1$, and even $z \gg 1$) distance indicator data. Instead, we opt for ``$y$-redshift", defined as $y = [z/(1+z)]$, with reverse transformation $z = [y/(1-y)]$. (We provide arguments for the wholesale replacement of $z$-redshift with $y$-redshift within theoretical cosmology in Appendix~\ref{AppRedshift}.)

With these redshift variable definitions, one easily obtains: $(1 + z) = (1-y)^{-1}$, and: $dz = (1-y)^{-2} ~ dy$. Thus we may re-write Eq.~\ref{EqndLumVSz} as:
\begin{equation}
	d_{L} (y) = c ~ \frac{1}{(1 - y)} ~ \int^{y}_{0} \frac{1}{(1 - y')^{2}} ~ \frac{1}{H(y')} dy' ~.
	\label{EqndLumVSy}
\end{equation}
This expression may be \textit{inverted} to express $H(y)$ in terms of the Luminosity Distance -- where, as usual, the latter is obtained from the Distance Modulus $\mu$ data, via: $d_{L} [\textrm{Mpc}] = 10^{(\mu - 25)/5}$. We thus invert Eq.~\ref{EqndLumVSy} to obtain:
\begin{equation}
	H (y) = \frac{c}{(1 - y)^2} ~ \{ \frac{d}{dy} [(1 - y) * d_{L} (y)] \} ^{-1} ~.
	\label{EqnHyVSdLum}
\end{equation}

Let us define a ``fitting function":
\begin{equation}
	F_{fit} (y) \equiv [(1 - y) * d_{L} (y)] ~,
	\label{EqnFitFnDefn}
\end{equation}
and convert each SNIa data point from a distance modulus $\mu _{SN}$ value to its corresponding value of $F_{fit} (y_{SN})$. (The error covariance matrix entries are also transformed appropriately, using the $y$-redshifts of both of the SNIa in each covariant pair.) We then divide the full SNIa data set into separate $y$-redshift bins (of width $\Delta y$) that are small enough for $F_{fit} (y)$ in each bin to be sufficiently well modeled using a straight-line segment; and thus the \textit{slopes} of those line segments can be used to approximate the derivative $d/dy \{ F_{fit} (y) \}$ from Eq.~\ref{EqnHyVSdLum}, for each bin\footnote{Fitting a line to multiple SNIa in a redshift bin is a way of ``smoothing" the data that represents an alternative method (perhaps with less ``processing" of the data), to the Gaussian smoothing process described in \citet{ShafSmooth}, and demonstrated for the original Pantheon data \citep{PantheonOrigData} in \citet[e.g.,][]{SmoothShaf}.}. In other words, if the $i^{th}$ bin is fit with the line segment, $F_{i}(y) = (m_{i} \cdot y) + b_{i}$, then given the set of best-fit slopes $\{ m_{i} \}$ for all of the bins, we obtain the set of Hubble Parameter estimates for the bins as:
\begin{equation}
	\{ H_{i} \} = \{ \frac{c}{(1 - \bar{y}_{i})^2} ~ ( m_{i} ) ^{-1} \} ~,
	\label{EqnBinnedHiVals}
\end{equation}
where $\bar{y}_{i}$ is the ``inverse-variance-weighted-average" redshift for each bin, given by a sum over the ($N_{i}$) supernovae in the $i^{th}$ bin as:
\begin{equation}
	\bar{y}_{i} = [\sum^{N_{i}}_{k=1} (\frac{y_{k}}{\sigma _{k}^{2}}) ~ ] ~ / ~  [ \sum^{N_{i}}_{k=1} (\frac{1}{\sigma _{k}^{2}}) ~ ] ~.
	\label{EqnBinWtdAvgY}
\end{equation}
(These $\sigma _{k}$ are not the \textit{redshift} uncertainties -- which happen to not be available for all of the SNIa in our study -- but are rather the uncertainties in the $F_{fit} (y)$ value for each SNIa, meaning that the supernovae with the more uncertainly measured distance moduli are relatively de-weighted in computing  $\bar{y}_{i}$.)

Note that each bin is fitted \textit{independently} -- correlated error terms in the covariance matrix have been dropped between SNIa data points that found themselves sorted into different $y$-redshift bins. Additionally, we do \textit{not} require the line segments to meet continuously at the bin edges. Trying that initially, we found that it unwisely forces ``good" bins into alignment with neighboring ``bad" bins (i.e., bins that are data-poor, and/or have poorly measured data points). Instead, we find it more useful to maintain bin-independence, and simply regard bins with more uncertain fits as being less definitive than bins with higher quality data and better fits. (And more fundamentally, our goal is to approximate the tangent curve to the $F_{fit} (y)$ function; not to re-imagine $F_{fit} (y)$ as a piecewise-continuous function.)

\subsection{The Binning Requirements\label{BinningReqs}}

The bin-widths $\Delta y$ should not be arbitrarily chosen, but must properly obey both maximal and minimal $\Delta y$ restrictions. First we consider the maximum justifiable bin widths.

Given our methodology of ``Linear Cosmography", in which we fit straight line segments to each of the bins, we must guarantee for a Taylor expansion fit to the data, that the coefficients of the quadratic (and higher) components of the expansion should be very small compared to the linear coefficient. While we cannot assume a particular theoretical model for the detailed cosmic evolution here, for the purposes of approximately quantifying such a Taylor expansion, we can take advantage of the fact that standard $\Lambda$CDM is a ``fairly good" fit to the data.

Using the Friedmann equation \citep{KolbTurner}, and assuming spatial flatness, we have $H(y)$ being proportional to the square root of the total density, $\sqrt{\rho_{tot}}$. Neglecting radiation (and relativistic matter) for the late universe, and with current matter contribution $\Omega_{M}$, so that $(\Omega_{M} + \Omega_{\Lambda}) \simeq 1$, we thus have:
\begin{equation}
	H_{\Lambda}(y)/H_{0} \simeq \sqrt{[ \Omega_{M} \cdot (1-y)^{-3} ]  + (1 - \Omega_{M})} ~.
	\label{EqnHyForLCDM}
\end{equation}
We then utilize Eq.~\ref{EqnHyVSdLum} as follows:
\begin{equation}
	\frac{d}{dy} [F_{fit} (y)] \rightarrow \frac{c}{(1 - y)^2~H_{\Lambda} (y)} \equiv \frac{d}{dy} [F_{Tay}] ~,
	\label{EqnFitFnVsHy}
\end{equation}
where \textit{here} we are assuming the use of $H(y) \equiv H_{\Lambda}(y)$ from $\Lambda$CDM (i.e., Eq.~\ref{EqnHyForLCDM}) for the definition of the ``Taylor expansion function", $F_{Tay}$. (Note that the quadratic expansion term, coming from the second $y$-derivative of $F_{Tay}$, would be derived from the \textit{first} $y$-derivative of $\{c/[(1 - y)^2~H_{\Lambda} (y)]\}$.)

Thus defining: 
\begin{equation}
D_{N} \equiv \frac{d^{N}}{dy^{N}} \{ F_{Tay} \} |_{y=y_{0}} = \frac{d^{N-1}}{dy^{N-1}} \{ \frac{c}{(1 - y)^2~H_{\Lambda} (y)} \} |_{y=y_{0}} ~,
	\label{EqnDerivsOfFTay}
\end{equation}
and expanding around some central bin redshift value $y_{0}$, using the maximum range for the expansion as half of the bin width -- i.e., $|y - y_{0}| \leq (\Delta y/2$) -- we then have, at the bin edges:
\begin{equation}
	F_{Tay} (y) \simeq F_{Tay} (y_{0}) + [\frac{D_{1}}{1!} \cdot (\frac{\pm \Delta y}{2})] + [\frac{D_{2}}{2!} \cdot (\frac{\pm \Delta y}{2})^{2}] + \dots ~.
	\label{EqnExpandingFTay}
\end{equation}
If our assumption is to be valid that \textit{most} of the dynamics of $F_{Tay}$ within a bin is due to the linear variation, then the quadratic term must be much smaller than the linear term. (The higher terms will be even smaller, since $\Delta y$ here is quite small.) Setting our criterion for ``smallness" as demanding that the quadratic term be no bigger than $\sim 1 \%$ of the linear term, we thus get the requirement: 
\begin{equation}
	\Delta y \lesssim (0.01) \cdot 4 ~ |D_{1}/D_{2}| ~.
	\label{EqnDeltaYreq}
\end{equation}
Numerically, this depends upon the $\Lambda$CDM model parameters used, and the central redshift value $y_{0}$ of the particular bin under consideration. From ACT DR6 \citep{ACTDR6} we take $\Omega_{M} = 0.3032$. As we will see, the redshifts of the SNIa used in this analysis cover the range $y \sim 0.0148 - 0.5858$ (i.e., $z \sim 0.015 - 1.414$). Using these numerical values in Eq.~\ref{EqnDeltaYreq}, the requirement at the low-redshift end of the data is $\Delta y \lesssim 0.0257$, and at the high-redshift end is $\Delta y \lesssim 0.0233$. Since we employ equal-width y-redshift binnings in this paper, we have chosen to simplify things by implementing the maximum bin width requirement: $\Delta y \leq 0.024$.

Alternatively, the bin widths must not be \textit{too small}, or several bins will have too few SNIa to obtain useful estimates of the $\{ H_{i} \}$ values. Furthermore, observational redshift uncertainties $\sigma_{y}$ would lead to substantially wrong placements of SNIa within the bins, and (even worse) may lead to a number of SNIa being assigned into the wrong bins entirely. This issue is sensitive to the specific data sets being used for the study, which we discuss in more detail below. But here we note that we have found it pragmatically necessary to implement the minimum bin width requirement: $\Delta y \geq 0.012$.

Combining these requirements, we have conducted runs with equal-spaced binning choices of: $\Delta y = \{ 0.012, 0.015, 0.018, 0.021, 0.024 \}$.

Lastly, we note that we could \textit{start} the binning at $y_{init} = 0$, or just below the smallest-redshift SNIa in the data samples, or anywhere in between. We have tried variations of the above choices, and have found that skipping the ``empty redshift space" and starting the binning just below where the data begins is best for ensuring a more equal sharing of SNIa within the first couple of bins. Thus we typically begin the binning somewhere around $y_{init} \simeq 0.014$.

\subsection{The SNIa Data Sets Used in This Study \label{SecSNIaData}}

In our current era of abundant SNIa compilations to choose from, it is not always clear which data set (or data sets) to use. For example, in the DESI DR2 BAO study \citep{DESIBAO}, they obtain three disjoint error ellipses for constraining dark energy parameters, by performing separate analyses combining their BAO data one-by-one with three of the major SNIa compilations: Pantheon+ \citep{PantheonPlusScol}, the Dark Energy Survey 5-Year (``DES5Yr") data release \citep{DES5YrSNIa}, and the Union3 compilation \citep{Union3Rubin}. In this paper, we must determine which (combination of) data sets to use, complicated by the fact that none of them are truly optimized for the analysis that we are attempting, and each have internal limitations (such as gaps at crucial redshift epochs in cosmic history).

First, considering DES5Yr versus Pantheon+, we opt for DES5Yr as our main SNIa data set. Besides being a more recent compilation, DES5Yr\footnote{DES5Yr SNIa data were formerly available online at:\\ https://github.com/des-science/DES-SN5YR/tree/main/4$\_$DISTANCES$\_$COVMAT; though this link has since been re-purposed for the recent DES-Dovekie update. DES5Yr zipped data can currently be downloaded from the \citet{DESnoSystematics} Legacy Release website: https://github.com/des-science/DES-SN5YR/releases/tag/1.3} contains a much larger SNIa sample at higher redshifts (taken from a ``homogeneous sample of high quality, well-calibrated light curves"), and is specifically designed to minimize systematics (even at the cost of sacrificing some statistical constraining power) by keeping only the highest quality SNIa, both at low and high redshifts. 

A problem for us with DES5Yr, however, is that there is a substantial \textit{gap} with very sparse data in the general vicinity of $y \sim z \sim 0.1$ (plus other smaller but significant gaps in places throughout the redshift range). This is mainly due to the fact that the DES5Yr compilation is a ``knitting together" of two separate data sets: the DES-observed sample proper, and a compilation of external, low-redshift data sets. As we will show, we find a great deal of ``activity" in $H(y)$ right around that deep gap, so that the lack of data there makes it very difficult to study these effects reliably. As has been much discussed recently \citep[e.g.,][]{PreventsLateH0SOlv}, there is an apparent ``crossing of the phantom divide" (dark energy equation of state $w < -1$) in the DESI BAO data \citep{DESIBAO} at intermediate redshift -- that is, around $z \sim 0.4$ ($y \sim 0.285$) or so -- followed by a possible ``thawing" (weakening dark energy) at later times, as $z \rightarrow 0$ \citep{DES5YrSNIa}. Such effects (particularly the phantom crossing) are difficult to reconcile with traditional dark energy models obeying the Dominant Energy Condition \citep[e.g.,][]{HawkingEllis}. We see hints of similar activity, but such data gaps in DES5Yr frustrate our attempts to clearly measure it.

(It is of course suspicious when one detects ``interesting" physics right at the juncture where two data sets are fused together; though in a detailed study by \citet{DESnoSystematics}, they claim: ``Here we present extensive modeling of this combined sample and validate the entire analysis pipeline used to derive distances... In all cases, statistical uncertainties dominate over systematics.")

In an earlier iteration of our work \citep{BochnerGR24}, uneven (``bespoke") bins were created for working around the DES5Yr gaps; essentially ``gerrymandering" the data set, which is not good practice, since it may lead to unpredictable biasing of the results. In this paper, we prefer to create more regular binning schemes (and in particular, evenly-spaced bins), which requires us to supplement DES5Yr with an additional data set. This, however, creates difficult choices.

First, it does not make sense to combine Pantheon+ with DES5Yr, since the latter is in part extracted from a re-analysis of the former; we find (based on SNIa names) that the two data sets appear to have $\sim$335 SNIa in common (over $18 \%$ of the DES5Yr sample). Also (as stated above), the choices of data cuts are done with different aims -- such as exchanging statistical power for lower systematics with DES5Yr. Furthermore, Pantheon+ has a similar data gap in the same place (around $y \sim z \sim 0.1$), so this combination would not be particularly helpful for solving the problem at hand.

Our first choice would be to combine (or perhaps replace) DES5Yr with the Union3 SNIa compilation, containing well over 2000 SNIa. This is impossible, however, since Union3 has not released a full data set containing distance modulus information (and supporting data) for each individual supernova, as had been done for previous SNIa compilations -- such as the Supernova Cosmology Project (SCP) Union2.1 compilation. In contrast, Union3 has only released frequentist spline-interpolated SN distances (which are not obviously amenable for our choices of re-binning); and cosmological theory tests with their Union3 data set are intended to be done within their ``UNITY1.5" \citep{Union3Rubin} software environment. We have inquired personally\footnote{D. Rubin, personal communication.}, and have been informed that: (i) Transformations such as our Eq.~\ref{EqnHyVSdLum} (and the subsequent discussion) cannot be performed within the UNITY environment; (ii) It is not yet possible with UNITY to combine Union3 with other SNIa compilations; and lastly, (iii) ``...there is no Union3 mu(z) outside of UNITY"\footnote{See also: https://astronomy.stackexchange.com/questions/59441/where-to-publicly-access-full-union-3-sn-dataset}. Thus we cannot utilize Union3 data for our analysis.

\begin{figure}[h!]
	\centering
	\epsscale{0.50}
	\plotone{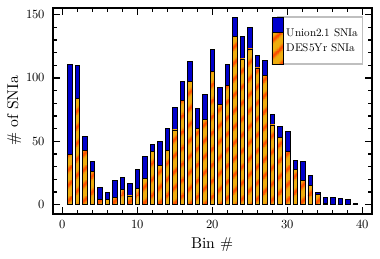}
	\caption{Population of each bin (for trial binning with $\{ y_{init}, \Delta y \} = \{ 0.014, 0.015 \}$), versus SNIa data set. The histogram demonstrates how this combination, adding in the Union2.1 compilation (with duplicated SNIa removed), helps fill in gaps in DES5Yr redshift-space data coverage. (The bins cover the redshift range from $y \sim z \sim 0.014$ to $y \sim 0.6$ ($z \sim 1.5$).) \label{FigDesVsUnionSNeBins}}
\end{figure}

The best remaining option for helping to fill the DES5Yr gaps in redshift coverage, is to combine it with the older (but still very useful) Union2.1 SNIa compilation \citep{SCPUnion21}. While Union2.1\footnote{Union2.1 SNIa data are available online at: https://supernova.lbl.gov/union} contains only about $\sim30\%$ as much data as DES5Yr, it extends both higher and lower in redshift distribution, and is more evenly spaced (lacking significant gaps). Union2.1 helps to (partially) fill in the gaps from DES5Yr, making for a smoother result for our series of bins. An example of how the two data compilations comparatively populate the bins is shown for the $\{ y_{init}, \Delta y \} = \{ 0.014, 0.015 \}$ case in Figure~\ref{FigDesVsUnionSNeBins}, where it is obvious how important the Union2.1 SNIa are for supplementing the sparse DES5Yr data within a gap extending approximately from $y \sim 0.08-0.16$ (i.e., $z \sim 0.09-0.19$).

\begin{figure}[h!]
	\centering
	\epsscale{0.50}
	\plotone{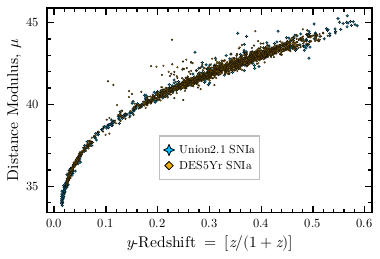}
	\caption{Distance Modulus $\mu$ ($= 5 Log_{10}[d_{L}] + 25$), versus ``$y$-Redshift", for our total data set produced by combining all data from the DES5Yr and Union2.1 SNIa compilations, except for the duplicated SNIa having been removed from the Union2.1 data set. (All points shown without error bars, for clarity.) \label{FigDesPlusUnionSNe}}
\end{figure}

This combination of data sets is not without its challenges. First, there are again SNIa in common between the two compilations, but now it appears to total just 49 duplicated SNIa (only $\sim8\%$ of the Union2.1 data set, and $< 3\%$ of the DES5Yr data set). Rather than attempt to reconcile the DES versus Union data differences for these common SNIa, we instead simply drop the overlapping SNIa from the Union2.1 set, choosing to defer to the more modern DES5Yr standardization of their compilation. Thus our final data set combination includes a total of $1829 + (580 - 49) = 2360$ distinct SNIa. A plot of distance modulus $\mu$ versus $y$-redshift of the combined data set is shown in Figure~\ref{FigDesPlusUnionSNe}. (Interestingly, note that many of the worst outliers -- particularly at lower redshift -- are actually from the DES5Yr compilation, rather than from the much older Union2.1 compilation.)

Furthermore, there will obviously be different systematics (and differences in the handling of them) between DES5Yr and Union2.1, which we can only characterize or ameliorate to a limited degree. For example, for the 49 duplicated SNIa between the two data sets, each data point has different values of distance modulus $\mu$ \textit{and} redshift $z$ between the two compilations. In DES5Yr, the distance moduli are \textit{higher} by a median value of $\Delta \mu = 0.043$ -- compare this to the effective uncertainty value 0.157, which is the median of the $\sigma_{\mu}$ values (i.e., the square-roots of the diagonal covariance matrix values), averaged between DES5Yr and Union2.1. (Incidentally, the DES5Yr redshifts are also higher, by a median value of $\Delta y = 0.0006$, which can be compared to the median of the $\sigma_{y}$ values from DES5Yr of 0.0008; and noting that Union2.1 does not appear to provide redshift uncertainty numbers.)

This offset in $\mu$ is suspiciously close to the ``$\Delta \mu_{\textrm{offset}}$" value of 0.042 quoted in \citet{DESnoSystematics} for duplicated SNIa between the DES5Yr and Pantheon+ compilations -- due apparently to updated methods of supernova fitting, host parameter estimates, and bias handling -- implying that Union2.1 is roughly in line with Pantheon+. It would therefore be tempting to try to ``correct" the Union2.1 compilation by adding $\Delta \mu \sim 0.042$ to each of the SNIa distance modulus values before combining the two data sets. However, this would not be appropriate, because this $\Delta \mu_{\textrm{offset}}$ is not a \textit{universal} offset between all of the supernovae in DES5Yr versus Pantheon+, but merely a \textit{difference} between whatever offsets exist in low-redshift versus high-redshift SNIa \citep{DESnoSystematics,EfStathPanthOff} -- with much of the offset implied to be due to the low-redshift, ``Foundation" SNIa data inherited by DES5Yr\footnote{Such a possible mismatch between low- versus high-redshift SNIa -- due to systematics or differences in SNIa modeling techniques -- is another indication of the benefits of our Hubble Tomography approach for \textit{pinpointing} any deviations from the standard cosmology, rather than comparing models which are integrated over the entire redshift range of the mixed data sets.}. (In fact, all of our duplicated SNIa, with this notable offset, are confined to low redshifts: $z \sim y \lesssim 0.07$.) Because the complexity of undertaking a complete re-analysis of the Union2.1 SNIa data set, using the DES5Yr fitting methodology, is beyond the scope of this paper, there is no way of knowing if this general distance modulus offset for these few low-redshift ``duplicated" SNIa is characteristic of the rest of the DES5Yr and Union2.1 data sets. We must therefore accept this potential offset (most notably at low redshifts) between our two SNIa data samples as a known but likely small systematic issue. (As a test, we \textit{did} conduct simulations with this $\Delta \mu \equiv 0.042$ offset being experimentally added in to the entire Union2.1 data set, and found that while the fit results do get moved around a bit by the adjustment, the overall visual impressions of the results do not change in any meaningful way.)

Other differences between the data sets includes a different treatment of errors: such the aforementioned fact that DES5Yr provides redshift uncertainties (as $\sigma_{z}$ values), while Union2.1 does not; and also the unusual procedure by DES5Yr of folding uncertainties about the SNIa classifications into the $\sigma_{\mu}$ numbers, instead of just specifying a hard cut on supernovae with uncertain classification as Type Ia. 

Since there is no panacea for these data set difficulties -- not until a more complete compilation is available to us (particularly in the $y \sim z \sim 0.1$ range) -- we go ahead with this analysis, while recognizing that we cannot quantify our findings yet with statistical rigor. (Hence we will not refer to our findings as ``proof" or ``evidence" or even ``signs", but rather as ``hints".) We do so, on the one hand, to demonstrate our method of Hubble Tomography via Linear Cosmography; and also, because -- much as in the original discoveries of both the Hubble Expansion, and of its acceleration -- interesting qualitative behaviors can sometimes be discovered, even when the statistical power of the findings may not (at least in retrospect) be considered certain.

\subsection{The Method of Best-Fitting for Binned Data \label{SecFittingBins}}

\begin{figure}[h!]
	\centering
	\epsscale{0.50}
	\plotone{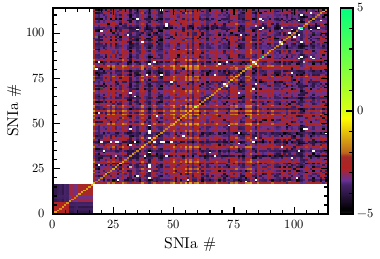}
	\caption{The (statistical + systematics) $\mu$-error covariance matrix ($Log_{10}$ scale) for Bin \#17 of the $\{ y_{init}, \Delta y \} = \{ 0.014, 0.015 \}$ case, covering the $y$-redshift range $y \in [0.254,0.269]$. The matrix is block diagonal -- covariances between the (entirely distinct) SNIa of the two different compilations are set to zero -- where the lower-left block represents Union2.1 SNIa, while the upper-right block represents DES5Yr data. (The numbers along the axes represent the arbitrary \# of each supernova in the bin.)
	\label{FigBinCovMat}}
\end{figure}

In preparation for linear fitting, the full error covariance matrices for each of the two compilations (DES5Yr and Union2.1) are reduced to block diagonal matrices (one block for each redshift bin), by dropping covariances between SNIa placed into different bins. Then these two matrices are merged in such a way so that the covariance sub-matrix for each redshift bin of the summed sample is now itself made into a block diagonal (with 2 distinct blocks), by adding together the pertinent (DES5Yr + Union) data, without assuming any covariance values between SNIa from the different compilations. (It would be of questionable utility to try to link the data sets by using covariances from the few duplicated SNIa.) A sample covariance matrix of a ``typical" bin in shown in Figure~\ref{FigBinCovMat}.

We then perform the optimization using standard statistical methods, similar to the method\footnote{This method was originally adapted from ``Chi-Square Minimization" by John W. Fowler, 2008,\\ https://web.ipac.caltech.edu/staff/fmasci/home/astro\_refs/Chi-square-min.pdf} from our previous cosmographic study \citep{BochnerLamDev}, now including more complete SNIa covariance information, though here only using 2 cosmographic fitting coefficients: i.e., $F(y) \equiv p_{0} + p_{1} y$. Each SNIa data point has redshift $y_{j}$, and its distance  modulus $\mu_{j}$ is turned into a ``fitting function" value $f_{j}$, as per $F_{fit} (y)$ defined in Eq.~\ref{EqnFitFnDefn}. For each bin, its error covariance sub-matrix is similarly transformed into a (fitting-function) error matrix $\mathbf{\Omega_{F}}$, via:
\begin{equation}
	\mathbf{\Omega_{F}} = [\Omega_{F}]_{j,k} \equiv [ \sigma_{F_{fit}} ] _{j,k} = \frac{\partial [F_{fit}]}{\partial \mu} |_{y=y_{j}} \cdot \frac{\partial [F_{fit}]}{\partial \mu} |_{y=y_{k}} \cdot [ \sigma_\mu ]_{j,k} ~.
	\label{EqnCovTransform}
\end{equation}
This $\mathbf{\Omega_{F}}$ is then inverted to form the ``weight matrix", $\textbf{\textit{W}} \equiv \{ W_{jk} \} = \mathbf{\Omega_{F}} ^{-1}$. We use all of this to calculate the elements for the $2 \times 2$ coefficient matrix $\textbf{\textit{A}} \equiv \{ A_{mn} \}$, and an auxiliary vector $\textbf{\textit{B}} \equiv \{ B_{n} \} $, as follows (where $N_{i}$ is the number of SNIa in the $i^{th}$ bin):
\begin{eqnarray}
	 A_{00} = \sum^{N_{i}}_{j = 1} \sum^{N_{i}}_{k = 1} W_{jk} ~,~~~
	 A_{10} = A_{01} & = & \frac{1}{2} \sum^{N_{i}}_{j = 1} \sum^{N_{i}}_{k = 1} W_{jk} (y_{j} + y_{k}) ~,~~~
	 A_{11} = \sum^{N_{i}}_{j = 1} \sum^{N_{i}}_{k = 1} W_{jk} (y_{j} \cdot y_{k}) ~, 
	 \\ 
	 B_{0} = \frac{1}{2} \sum^{N_{i}}_{j = 1} \sum^{N_{i}}_{k = 1} W_{jk} (f_{j} + f_{k}) ~&,&~~~ 
	 B_{1} = \frac{1}{2} \sum^{N_{i}}_{j = 1} \sum^{N_{i}}_{k = 1} W_{jk} (f_{j} \cdot y_{k} + f_{k} \cdot y_{j}) ~.
	\label{EqnFisherMatrix2x2}
\end{eqnarray}
Finally, the best-fit line segment parameters for the bin are obtained via: $\textbf{\textit{p}} \equiv \{p_{0},p_{1}\} = \textbf{\textit{A}} ^{-1} \textbf{\textit{B}}$, with the error covariance matrix for these parameters $\textbf{\textit{p}}$ being obtained as \textit{\textbf{E}}~$\equiv \textbf{\textit{A}} ^{-1}$. (These $\{p_{1,i}\}$ for each of the bins correspond to the set of slopes, $\{m_{i}\}$, in Eq.~\ref{EqnBinnedHiVals}; with slope uncertainties $\{ \sigma_{m,i} \}$ being $\{ (E_{11,i} )^{1/2} \}$.)

To simplify the analysis, we do not directly incorporate redshift uncertainties into the line-fitting process. (This should be acceptable, since we are not calculating formal error ellipses for the cosmological parameter estimates.) It would be difficult to include them here, since these uncertainties are provided with the DES5Yr data set (as $z$-redshift uncertainties, $\sigma_{z}$), but do not appear to be available in the Union2.1 SNIa public data release. Yet we note that the DES5Yr $y$-redshift uncertainties, calculated via\footnote{Note that the strong suppression of $\sigma_{y}$ relative to $\sigma_{z}$ at high redshifts, by the factor $(1-y)^{2}$, is yet another reason in favor of working exclusively with $y$-redshift.}: $\sigma_{y} = \sigma_{z} / (1+z)^{2} = \sigma_{z} \cdot (1-y)^{2}$, are not insignificant, and create several concerns for our method. First, uncertainty causes misplacement of SNIa within their redshift bins, which affects the slope of the fitted lines (and hence the $H(y)$ estimates). But more perniciously, substantial $\sigma_{y}$ values can lead to a number of SNIa being erroneously placed into the wrong bins entirely, which complicates the process of even obtaining well-understood $\sigma_{H}$ values.

The typical size of $\sigma_{y}$ values in a data set clearly affects the minimum bin width, $\Delta y$, that we can reliably choose. With the DES5Yr SNIa data set (for which we have this information), and for our $\Delta y = 0.015$ binning choice (which we will highlight in the most detail, below), we note that all of the bins have median $\sigma_{y} < 0.00122$, which in all cases is $\lesssim 8\%$ of the bin widths $\Delta y$. (Though specifically, for bins $\#2$ and $\#5$, the \textit{average} -- and maximal -- values of $\sigma_{y}$ are moderately higher; but that is most likely due to a small number of poor SNIa data points, presumably with large $\sigma_{\mu}$ values, which would therefore be de-emphasized in the line-fitting process.) For our smallest binning size, $\Delta y = 0.012$, the bin-median $\sigma_{y}$ values still all remain $\lesssim 10\%$ of $\Delta y$.

While these redshift uncertainties are not comfortably negligible, we will regard the data as ``good enough" for the demonstration of the Hubble Tomography method in this paper. The torrent of new SNIa data expected during the next several years will obviously help a great deal with improving the precision and reliability of the results from this method; however, the much heavier use of photometric redshifts -- as opposed to spectroscopically-measured redshifts -- is particularly problematic for a redshift-binning method such as this one. Resorting to photometric redshifts (and the consequently large redshift uncertainties) is obviously necessary for the goal of obtaining extremely large data sets of faint, high-redshift SNIa. Even though high-redshift SNIa are not our main focus, with most of the physics of interest to us (as previously discussed) happening after the acceleration epoch has begun ($y \lesssim 0.4$, $z \lesssim 0.66$), there is the much more serious problem of what may be called ``redshift aliasing", where high-redshift SNIa are erroneously classified as low-redshift SNIa, and would thus be improperly sorted into those crucial bins. A study of the challenges with photometric redshift estimates is conducted in \citet{PhotozFromLSST}; and there the authors demonstrate their method of ``training sample augmentation" for helping to ameliorate such difficulties. We must hope that this method (and perhaps others) will help improve the reliability of our Hubble Tomography method, when applied to the vast windfall of expected future supernovae.

\subsection{Incorporating BAO Data for Transition-Redshift Information \label{SecBAOdata}}

This paper is not the only work that has been done on the binning of the Hubble Parameter; for example, in \citet{DragonH}, they construct a ``modified-H" model in a more standard way than we do here -- rather than directly fitting $H(y_{SN})$ data points, they integrate a perturbed cosmological model to get the distance functions needed for BAO comparisons, allowing discrete deviations of $H(z)$ to be defined within 6 bin ranges. Their analysis extends to relatively high redshift ($z \in [0,2.5]$, and $z \in [0,5]$), while their findings seem to show the ``interesting" deviations from $\Lambda$CDM to be mostly confined to $z$ well below 1. They choose their binning to roughly match the binning of the DESI BAO DR2 data (presumably to aid in the comparison with BAO information), but which has the side-effect of restricting their analysis to contain only 6 bins; with, most significantly, only 1 redshift bin representing the wide range $z \in [0,0.4]$ ($y \in [0,0.29]$), and only 2 bins to cover nearly the entire post-acceleration era, $z < 0.6$ ($y < 0.375$). As one can see in the bottom panel of their Figure 1, the SNIa data is a much finer tool than BAO for mapping out the Hubble expansion at lower redshifts ($z \lesssim 1$), and we would like to use it as such.

First, to quantify the redshift of the deceleration-to-acceleration cosmic transition, we consider the Friedmann acceleration equation \citep{KolbTurner}, in the context of the $\Lambda$CDM model (treating it as a good-enough placeholder for the true cosmic evolution up to that time). In terms of the closure density fractions (at $t = t_{0}$) of matter ($\Omega_{M}$) and vacuum energy ($\Omega_{\Lambda}$), we have:
\begin{equation}
	\ddot{a} \propto -(\rho_{Tot} + 3 P_{Tot}) \propto - (\Omega_{M} - 2 \Omega_{\Lambda}) ~,
	\label{EqnFriedAccel}
\end{equation}
since the pressures of each component are $P_{M} = 0$, $P_{\Lambda} = - \rho_{\Lambda} \propto - \Omega_{\Lambda}$. Given that $\rho_{M} \propto a^{-3}$, and $\rho_{\Lambda} \propto a^{0}$, the instantaneous acceleration evolves as: $\ddot{a} \propto - \{ [\Omega_{M} \cdot (a/a_{0})^{-3}] - 2 \Omega_{\Lambda} \}$. The transition occurs at the moment of coasting, $\ddot{a} = 0$, with the transition scale factor, $y$-redshift, and $z$-redshift given (respectively) by:
\begin{equation}
	a_{Tr} =  a_{0} \cdot [\frac{\Omega_{M}}{2\Omega_{\Lambda}}]^{1/3} ~,~ y_{Tr} = 1 -  [\frac{\Omega_{M}}{2\Omega_{\Lambda}}]^{1/3} ~,~ z_{Tr} = [\frac{2\Omega_{\Lambda}}{\Omega_{M}}]^{1/3} - 1 ~.
	\label{EqnAccelTrans}
\end{equation}
The time of this $\Lambda$CDM-approximated transition epoch then depends upon which set of cosmic parameter estimates one uses. Assuming flatness ($\Omega_{\Lambda} = 1 - \Omega_{M}$), and using results from Planck \citep[from Table 1 of][]{PlanckH0}, specifically $\Omega_{M} = 0.3147$, we get: $y_{Tr} = 0.3877$, $z_{Tr} = 0.6331$. Alternatively, using the recent results from ACT DR6 \citep[the ``P-ACT-LB" case of][]{ACTDR6}, specifically $\Omega_{M} = 0.3032$, we get: $y_{Tr} = 0.3986$, $z_{Tr} = 0.6627$. This roughly rounds to $y_{Tr} \approx 0.4$, $z_{Tr} \approx 2/3 \approx 0.66$, though we will precisely depict both estimated transition epochs in our figures below.

In order to thoroughly measure the behavior of the Hubble Parameter both before and after this transition epoch, we therefore wish to plot the DESI DR2 BAO results along with our SNIa linear cosmography fits, a task which is not difficult. From \citet{DESIBAO}, there are values of the BAO comoving distance measured in the transverse direction, $D_{M}(z)$; comoving distance measurements along the line-of-sight direction, $D_{H}(z)$; and a combination of the two, interpreted as the ``isotropic" comoving distance, $D_{V}(z)$. Now, $D_{M}(z)$ is an \textit{integrated} measure over redshift, hence it mixes together the behaviors at different epochs; and so $D_{M}(z)$ (and thus $D_{V}(z)$) go against the spirit of what we are trying to do here, which is to provide snapshots of the \textit{instantaneous} evolution of the Hubble Parameter. Alternatively, the line-of-sight direction measurements satisfy the simple relation, $D_{H}(z) = c/H(z)$ (i.e., $D_{H}(y) = c/H(y)$), hence it is perfect for our use.

Their Table IV lists $(D_{H}/r_{d})$ values good for Hubble Tomography at six different redshifts\footnote{Unfortunately, the DESI lowest-redshift (bright galaxy sample, ``BGS") measurement, at $z_{eff} = 0.295$, provides no $D_{H}$ numbers.}, where $r_{d}$ is the scale of the pre-recombination sound horizon. They measure the product, $(h \cdot r_{d}) = 101.54 \pm 0.73$ Mpc, requiring one to assume a value of $H_{0}$. Now since our work here only makes \textit{relative} comparisons between data sets (and theory), and because both the DES5Yr and Union2.1 SNIa data sets are both normalized to $H_{0} \equiv 70$~km~sec$^{-1}$ Mpc$^{-1}$ (i.e., $h = 0.7$)\footnote{$H_{0}$ normalization noted online for Union 2.1 at: https://supernova.lbl.gov/union/figures/SCPUnion2.1$\_$mu$\_$vs$\_$z.txt; and noted in the README for the ``4\_DISTANCES\_COVMAT" section of the DES5Yr and DES-Dovekie data repositories we have specified earlier.}, it is most convenient to set $h = 0.7$ for these BAO data as well (and also for the $\Lambda$CDM theoretical model to be used for comparisons). Hence we convert the DESI DR2 BAO results to $H(y)$ values via:
\begin{equation}
	H(y)_{BAO} = c \cdot \frac{0.7}{101.54} \cdot \frac{1}{(D_{H}/r_{d})_{DESI}} ~,
	\label{EqnHubbFromBAO}
\end{equation}
(with the correspondingly appropriate transformation done for their uncertainties).

As will be evident in the results below, the BAO data -- as well as the SNIa data -- are seen to exhibit strong changes (relative to $\Lambda$CDM) \textit{right in the vicinity} of the transition redshift $y_{Tr}$, quite possibly indicating a type of ``phase transition" in the cosmic evolution just as the acceleration was beginning to take over the universe.

\section{Linear Cosmography Fitting Results \label{SecCosmogResults}}

We will display a comprehensive suite of results for our cosmographic model with $\{ y_{init}, \Delta y \} = \{ 0.014, 0.015 \}$, which we choose to be the base model for this demonstration, since the results are most evocative for this particular case. But we note again that the results are \textit{not} currently proven to be robust against binning scheme changes, so we will therefore also display a selection of results for the different cases specified above in Sec.~\ref{BinningReqs}.

\begin{figure*}[ht!]
	\begin{subfigure}{.475\linewidth}
		\includegraphics[width=\linewidth]{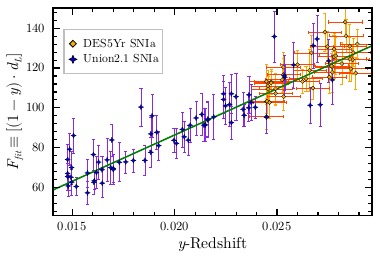}
		\caption{}
		\label{LineFitsBin1}
	\end{subfigure}\hfill
	~ 
	\begin{subfigure}{.475\linewidth}
		\includegraphics[width=\linewidth]{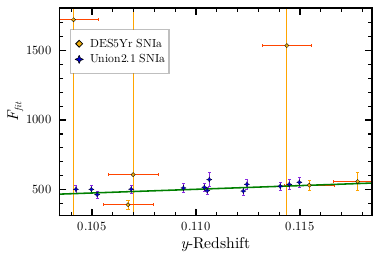}
		\caption{}
		\label{LineFitsBin7}
	\end{subfigure}
	\medskip
	\begin{subfigure}{.475\linewidth}
		\includegraphics[width=\linewidth]{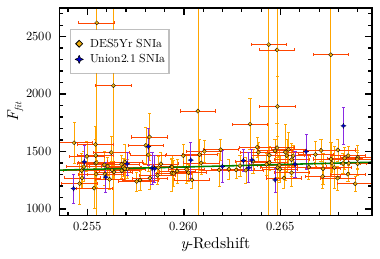}
		\caption{}
		\label{LineFitsBin17}
	\end{subfigure}\hfill
	\begin{subfigure}{.475\linewidth}
		\includegraphics[width=\linewidth]{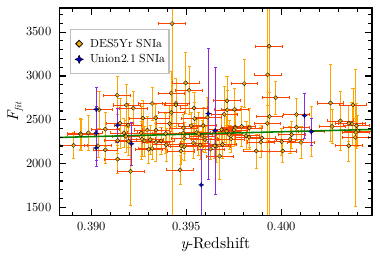}
		\caption{}
		\label{LineFitsBin26}
	\end{subfigure}
	\caption{Linear Cosmography best-fits (solid green lines) shown for four representative bins of interest for the $\{ y_{init}, \Delta y \} = \{ 0.014, 0.015 \}$ case: (a) Bin \#1 ($y \in [0.014,0.029]$), (b) Bin \#7 ($y \in [0.104,0.119]$), (c) Bin \#17 ($y \in [0.254,0.269]$), and (d) Bin \#26 ($y \in [0.389,0.404]$). Each plot depicts a section of our combined \{DES5Yr + Union2.1\} SNIa data set, showing the Fitting Function $F_{fit}$ (Eq.~\ref{EqnFitFnDefn}) versus $y$-redshift. In addition to the uncertainties in $F_{fit}$ (vertical error bars) plotted for all supernovae here, redshift uncertainties (horizontal error bars) are also available and shown for the DES5Yr SNIa.}
	\label{LineFitsFourBins}
\end{figure*}

First, Figure~\ref{LineFitsFourBins} displays a sample of linear cosmography fits for four representative bins, each of which has some property of interest. In order, for Sub-Figures~\ref{LineFitsFourBins}(a)-(d): Bin \#1 shows the beginning of the data at the low-redshift end, with Union2.1 data providing an anchor before the DES5Yr supernovae begin; Bin \#7 shows an example redshift range where sparse (and high-uncertainty) DES5Yr data are supplemented by several well-measured Union2.1 SNIa, so that a line fit with some confidence is possible; Bin \#17 shows a more ``typical" redshift range, where DES5Yr supernovae predominate, but Union2.1 data contribute as well; and Bin \#26 is located in the vicinity of the deceleration-to-acceleration transition, $y_{Tr} \approx 0.4$.

\begin{figure*}[ht!]
	\begin{subfigure}{.475\linewidth}
		\includegraphics[width=\linewidth]{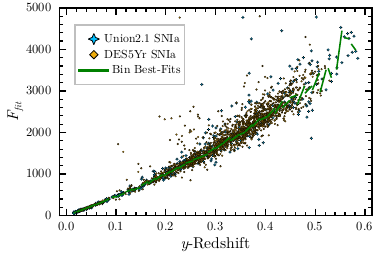}
		\caption{}
		\label{AllFitsFitFnPlot}
	\end{subfigure}\hfill
	~ 
	\begin{subfigure}{.475\linewidth}
		\includegraphics[width=\linewidth]{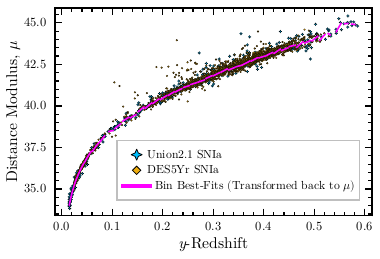}
		\caption{}
		\label{AllFitsMuPlot}
	\end{subfigure}
	\medskip
	\begin{subfigure}{.475\linewidth}
		\includegraphics[width=\linewidth]{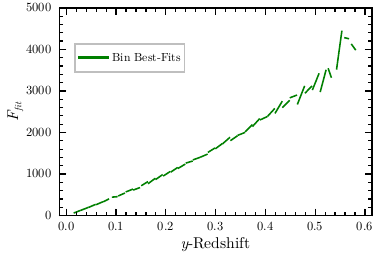}
		\caption{}
		\label{AllFitsFitFnNoDataPlot}
	\end{subfigure}\hfill
	\begin{subfigure}{.475\linewidth}
		\includegraphics[width=\linewidth]{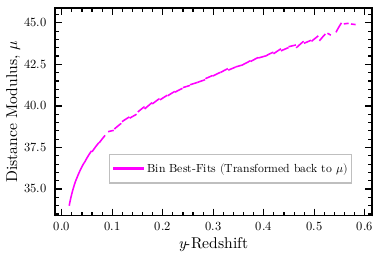}
		\caption{}
		\label{AllFitsMuNoDataPlot}
	\end{subfigure}
	\caption{Summary plots containing SNIa data and linear cosmography fits (short line segments) for all redshift bins. Plot (a) here depicts these with $F_{fit}$ as the vertical axis, while Plot (b) shows this axis transformed back into distance modulus $\mu$. Plots (c) and (d) show each of these, respectively, with the SNIa data omitted for visual clarity.}
	\label{LineFitsAllBins}
\end{figure*}

Next, Figure~\ref{LineFitsAllBins} shows all of the SNIa data, over the full redshift range, displaying all of the linear cosmography fits together in each of the plots. (Recall, as explained in Sec.~\ref{yVSzRedshift}, that we do \textit{not} require the independently-best-fit line segments to meet at the bin edges.) Fig.~\ref{AllFitsFitFnPlot} shows these with the Fitting Function $F_{fit} (y) = [(1 - y) * d_{L} (y)]$ as the $y$-axis of the plot (in which the fits are exactly straight line segments). Fig.~\ref{AllFitsMuPlot} shows the SNIa (and the fits) with the $y$-axis of the plot being transformed back into distance modulus $\mu$ (by reversing the transformation of Equation~\ref{EqnFitFnDefn}), so that the best-fit line segments are turned here into small sections of curved lines. Fig's.~\ref{LineFitsAllBins}c,d show the same as the above two figures, respectively, but with the SNIa data \textit{removed} from the plots, to produce an unobstructed view of the fitted cosmographic line segments themselves. These depictions of the combined fits provide a visual indication that some of the bins are likely not reliable -- especially those at the high-redshift end, but also with some questionable fits near the low-redshift data gap discussed in Section~\ref{SecSNIaData}, and at other scattered locations. Though overall, the line-segment fits appear to be a fairly good tracer of the supernova data. (Note that quantitative uncertainties on the resulting $H(y)$ values will be given below.)

\begin{figure*}[ht!]
	\begin{subfigure}{.475\linewidth}
		\includegraphics[width=\linewidth]{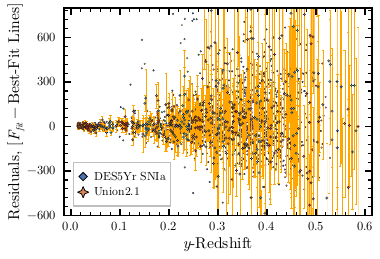}
		\caption{}
		\label{ResidsLineFitFnPlot}
	\end{subfigure}\hfill
	~ 
	\begin{subfigure}{.475\linewidth}
		\includegraphics[width=\linewidth]{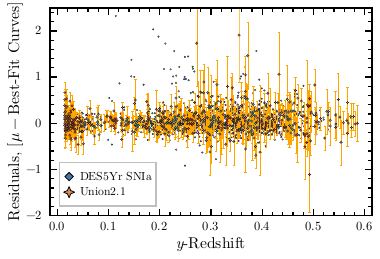}
		\caption{}
		\label{ResidsMuPlot}
	\end{subfigure}
	\caption{Plots (a) and (b) here show the fit residuals for (respectively) the best-fits presented above in Figures~\ref{LineFitsAllBins}a,b. Delineations between bins is indicated by using a different color for each distinct redshift bin of SNIa data points. Note that while the vertical error bars ($\sigma$'s) are included for the SCP Union2.1 data points, in the interest of visual clarity, the (often larger and much more numerous) $\sigma$ values are not shown here for the DES5Yr data points.}
	\label{TwoResidualsPlots}
\end{figure*}

Additionally, fit residuals corresponding to Figures~\ref{LineFitsAllBins}a,b (respectively) are shown in Figures~\ref{TwoResidualsPlots}a,b. While no obvious systematics resulting from this fitting method are immediately apparent, it is notable that the mathematical nature of $F_{fit} (y)$ causes their residuals to spread out more (though with appropriately larger error bars) at high redshift, relative to that at low redshift, than is the case for the (logarithmic) distance modulus $\mu$. While the $F_{fit}$ formalism is more applicable than fits to distance modulus for getting \textit{direct} estimates of the temporal evolution of Hubble Parameter $H(y)$, this effect does provide some additional challenge for this kind of Hubble Tomography.

The numerical Hubble Parameter results for our linear cosmography fits for all bins are given in Table~\ref{TableHyLinCosmogResults}. For comparison, $H(y)$ values at corresponding redshifts for a flat $\Lambda$CDM toy model with $H_{0} \equiv 70.0$ and $\Omega_{M} = (1 - \Omega_{\Lambda}) = 0.3032$ (from ACT DR6, \citet{ACTDR6}), are also given in the table; and presented along with these are the uncertainties on the Hubble Parameter estimates ($\sigma_{H(y)}$ values) for each bin, \textit{and} a calculation of ``how many sigmas away" each bin's $H(y)$ value is from that predicted by the $\Lambda$CDM model. Additionally, $H(y)$ values adapted from the DESI DR2 BAO results \citep{DESIBAO} are given in Table~\ref{TableHyBAOnumbers}.

\begin{deluxetable*}{ccccccc}
	\tablewidth{0pt}
	\tablecaption{Hubble Parameters from Binned Linear Cosmography Fits\label{TableHyLinCosmogResults}}
	\tablehead{
			\colhead{$\#$SNIa\tablenotemark{a}}
			& \colhead{$y$-Redshift\tablenotemark{b}} 
			& \colhead{$z$-Redshift\tablenotemark{c}} 
			& \colhead{$H(y)_{\Lambda CDM}$\tablenotemark{d,e}} 
			& \colhead{$H(y)_{Cosmog}$\tablenotemark{d,f}} 
			& \colhead{$\sigma_{H(y)}$\tablenotemark{d,f}}
			& \colhead{$N_{\Delta \sigma}$\tablenotemark{g}}
	}
	\startdata
	111 & 0.024 & 0.024 & 70.8 & 67.6 & 2.39 & -1.34 \\
	110 & 0.035 & 0.036 & 71.2 & 69.7 & 3.08 & -0.49 \\
	\phantom{0}54 & 0.050 & 0.053 & 71.7 & 64.6 & 4.89 & -1.45 \\
	\phantom{0}34 & 0.066 & 0.071 & 72.4 & 65.1 & 7.43 & -0.98 \\
	\phantom{0}14 & 0.081 & 0.088 & 73.0 & 58.0 & 16.9 & -0.89 \\
	\phantom{0}10 & 0.097 & 0.108 & 73.7 & 369 & 966 & +0.31 \\
	\phantom{0}19 & 0.110 & 0.124 & 74.3 & 71.6 & 27.1 & -0.10 \\
	\phantom{0}22 & 0.128 & 0.147 & 75.2 & 83.3 & 43.4 & +0.19 \\
	\phantom{0}17 & 0.141 & 0.164 & 75.9 & 92.6 & 42.5 & +0.39 \\
	\phantom{0}28 & 0.157 & 0.186 & 76.8 & 56.3 & 16.7 & -1.23 \\
	\phantom{0}38 & 0.173 & 0.210 & 77.7 & 49.6 & 11.2 & -2.51 \\
	\phantom{0}48 & 0.186 & 0.229 & 78.6 & 53.7 & 11.4 & -2.18 \\
	\phantom{0}50 & 0.202 & 0.254 & 79.6 & 60.5 & 17.0 & -1.12 \\
	\phantom{0}60 & 0.216 & 0.276 & 80.6 & 60.7 & 18.2 & -1.09 \\
	\phantom{0}77 & 0.232 & 0.302 & 81.8 & 71.4 & 19.7 & -0.53 \\
	\phantom{0}97 & 0.248 & 0.329 & 83.1 & 145 & 88.7 & +0.70 \\
	113 & 0.262 & 0.355 & 84.3 & 129 & 62.5 & +0.72 \\
	\phantom{0}76 & 0.276 & 0.382 & 85.6 & 112 & 68.9 & +0.38 \\
	\phantom{0}87 & 0.293 & 0.414 & 87.2 & 79.3 & 34.8 & -0.23 \\
	123 & 0.307 & 0.442 & 88.7 & 70.2 & 22.5 & -0.82 \\
	\phantom{0}93 & 0.321 & 0.472 & 90.3 & 61.5 & 22.0 & -1.31 \\
	111 & 0.336 & 0.507 & 92.2 & 81.9 & 32.8 & -0.31 \\
	148 & 0.351 & 0.542 & 94.1 & 150 & 89.5 & +0.62 \\
	133 & 0.366 & 0.578 & 96.2 & 62.5 & 18.3 & -1.84 \\
	140 & 0.381 & 0.616 & 98.4 & 59.6 & 17.8 & -2.18 \\
	118 & 0.396 & 0.655 & 100.8 & 148 & 121 & +0.39 \\
	114 & 0.410 & 0.695 & 103.2 & 65.8 & 22.5 & -1.66 \\
	\phantom{0}71 & 0.426 & 0.743 & 106.2 & 45.4 & 13.9 & -4.37 \\
	\phantom{0}62 & 0.442 & 0.791 & 109.3 & 81.8 & 49.2 & -0.56 \\
	\phantom{0}58 & 0.455 & 0.834 & 112.1 & 220 & 423 & +0.26 \\
	\phantom{0}35 & 0.471 & 0.890 & 115.9 & 35.2 & 13.8 & -5.85 \\
	\phantom{0}34 & 0.487 & 0.951 & 120.2 & 91.3 & 85.0 & -0.34 \\
	\phantom{0}23 & 0.502 & 1.009 & 124.4 & 40.5 & 17.7 & -4.74 \\
	\phantom{0}10 & 0.514 & 1.060 & 128.0 & 28.1 & 14.3 & -6.99 \\
	\phantom{0}6 & 0.529 & 1.125 & 132.9 & -39.5 & 92.1 & -1.87 \\
	\phantom{0}6 & 0.547 & 1.206 & 139.2 & 16.8 & 10.9 & -11.23 \\
	\phantom{0}5 & 0.565 & 1.299 & 146.5 & -565 & 1.2E+4 & -0.06 \\
	\phantom{0}4 & 0.577 & 1.361 & 151.6 & -99.6 & 410 & -0.61 \\
	\phantom{0}1 & 0.586 & 1.414 & 155.9 & N/A\tablenotemark{h} & N/A\tablenotemark{h} & N/A\tablenotemark{h} \\
	\enddata
	\tablenotetext{a}{The total number of supernovae from \{DES5Yr + Union2.1\} \textit{combined} for each bin.}
	\tablenotetext{b}{The value of $\bar{y}_{i}$ for each bin $i$, as per Eq.~\ref{EqnBinWtdAvgY}.}
	\tablenotetext{c}{Calculated via $\bar{z}_{i} \equiv \bar{y}_{i}/(1 - \bar{y}_{i})$ for each bin.}
	\tablenotetext{d}{All Hubble Parameter values and $\sigma$'s given in [$km~s^{-1}~Mpc^{-1}$].}
	\tablenotetext{e}{Calculated from a Flat $\Lambda$CDM model with $H_{0} = 70$, $\Omega_{M} = (1 - \Omega_{\Lambda}) = 0.3032$.}
	\tablenotetext{f}{Cosmography values and $\sigma$'s calculated via Eq's.~\ref{EqnBinnedHiVals}-\ref{EqnBinWtdAvgY}, and Sec.~\ref{SecFittingBins}.}
	\tablenotetext{g}{Calculated as $N_{\Delta \sigma} \equiv \{ [H(y)_{Cosmog} - H(y)_{\Lambda CDM}] / \sigma_{H(y)} \}$ \textit{(dimensionless)}.}
	\tablenotetext{h}{Since the final bin has a single supernova, $H(y)$ is undefined in our fitting method.}
\end{deluxetable*}

\begin{deluxetable*}{ccccccc}
	\tablewidth{0pt}
	\tablecaption{Hubble Parameter Values from DESI DR2 BAO Public Data \label{TableHyBAOnumbers}}
	\tablehead{
		\colhead{$y$-Redshift\tablenotemark{a}} 
		& \colhead{$z$-Redshift\tablenotemark{b}} 
		& \colhead{$H(y)_{\Lambda CDM}$\tablenotemark{c,d}} 
		& \colhead{$H(y)_{BAO}$\tablenotemark{c,e}} 
		& \colhead{$\sigma_{H(y)}$\tablenotemark{c,e}}
		& \colhead{$N_{\Delta \sigma}$\tablenotemark{f}}
		& \colhead{$\Delta_{H(y)}$\tablenotemark{c,g}}
	}
	\startdata
	0.338 &	0.510 &	\phantom{0}92.4 & \phantom{0}94.5 &	1.84 & +1.14 & +2.1 \\
	0.414 &	0.706 &	103.9 &	106.2 &	1.80 & +1.28 & +2.3 \\
	0.483 &	0.934 &	119.0 &	117.2 &	1.28 & -1.41 & -1.8 \\
	0.569 &	1.321 &	148.3 &	145.8 &	2.27 & -1.10 & -2.5 \\
	0.597 &	1.484 &	161.8 &	161.2 &	6.49 & -0.09 & -0.6 \\
	0.700 &	2.330 &	241.4 &	239.4 &	2.80 & -0.71 & -2.0 \\
	\enddata
	\tablenotetext{a}{Calculated via $y_{\textrm{eff}} \equiv z_{\textrm{eff}}/(1 + z_{\textrm{eff}})$ \{see footnote (b)\}.}
	\tablenotetext{b}{Taken from $z_{\textrm{eff}}$ in Table IV of \citet{DESIBAO}.}
	\tablenotetext{c}{All Hubble Parameter values and $\sigma$'s given in [$km~s^{-1}~Mpc^{-1}$].}
	\tablenotetext{d}{Calculated from Flat $\Lambda$CDM model with $H_{0} = 70$, $\Omega_{M} = (1 - \Omega_{\Lambda}) = 0.3032$.}
	\tablenotetext{e}{BAO values and $\sigma$'s calculated via Eq.~\ref{EqnHubbFromBAO}.}
	\tablenotetext{f}{Calculated as $N_{\Delta \sigma} \equiv \{ [H(y)_{BAO} - H(y)_{\Lambda CDM}] / \sigma_{H(y)} \}$ \textit{(dimensionless)}.}
	\tablenotetext{g}{Calculated as $\Delta_{H(y)} \equiv [H(y)_{BAO} - H(y)_{\Lambda CDM}]$ (expressed in [$km~s^{-1}~Mpc^{-1}$]).}
\end{deluxetable*}

One issue that is readily apparent from the analysis is the widely-varying quality of the fits, with $H(y)$ sometimes differing greatly (with very large $\sigma_{H(y)}$) from bin to bin; particularly at the ragged high-redshift end at the data, and also notably for Bin \#6, located right around the aforementioned data gap in DES 5Yr. This variability (and overly large deviations from the $\Lambda$CDM expectation) is in contrast to the BAO results, which vary much less, and remain closer to the $\Lambda$CDM model values.

Such variability in the cosmography results is obviously due, in part, to the comparatively high uncertainties and sparseness of the SNIa data in some of the bins; but some of it may be intrinsic to the data, since the BAO data are primarily located at higher redshift than are the supernovae -- back when the dark energy was less cosmologically dominant -- and the BAO analysis also averages the data together within bins much wider in redshift space (see Table III in \citet{DESIBAO}) than those in our SNIa study, potentially averaging away any natural variations which do exist.

\begin{figure*}[ht!]
	\begin{subfigure}{.475\linewidth}
		\includegraphics[width=\linewidth]{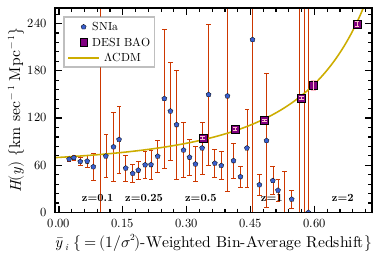}
		\caption{}
		\label{HofYMainPlot}
	\end{subfigure}\hfill
	~ 
	\begin{subfigure}{.475\linewidth}
		\includegraphics[width=\linewidth]{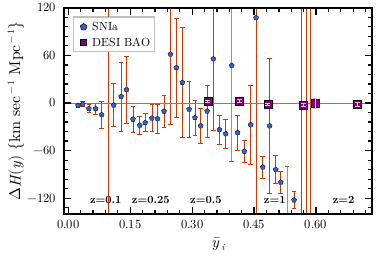}
		\caption{}
		\label{HofYResidsPlot}
	\end{subfigure}
	\caption{Hubble Parameter values, $H(y)$, derived from the linear cosmography fits in Fig.~\ref{LineFitsAllBins}, using our \{DES5Yr + Union2.1\} data set (light blue pentagons); shown with $H(y)$ values adapted from DESI DR2 BAO data (purple squares, most larger than their error bars). Plot (a) shows the $H(y)$ values, along with a flat $\Lambda$CDM toy model (solid orange line) with $H_{0} \equiv 70.0$ and $\Omega_{M} = (1 - \Omega_{\Lambda}) = 0.3032$. Plot (b) shows the residuals, [$H(y)_{Data} - H(y)_{\Lambda CDM}$], with the data the same as from Tables~\ref{TableHyLinCosmogResults},\ref{TableHyBAOnumbers}. (Each horizontal axis represents $\bar{y}_{i}$, the inverse-variance weighted average $y$-Redshift for each of the bins, as per Eq.~\ref{EqnBinWtdAvgY}.)}
	\label{TwoHofYplots}
\end{figure*}

A graphical view of the results in Tables~\ref{TableHyLinCosmogResults},\ref{TableHyBAOnumbers} is displayed in Figure~\ref{TwoHofYplots}, where Fig.~\ref{TwoHofYplots}a shows the $H(y)$ values from linear cosmography SNIa (and BAO) measurements, versus this $\Lambda$CDM toy model; and Fig.~\ref{TwoHofYplots}b shows the plot of $H(y)$ residuals, where the $\Lambda$CDM model values are subtracted from the cosmography and BAO data points.

\begin{figure*}[ht!]
	\begin{subfigure}{.475\linewidth}
		\includegraphics[width=\linewidth]{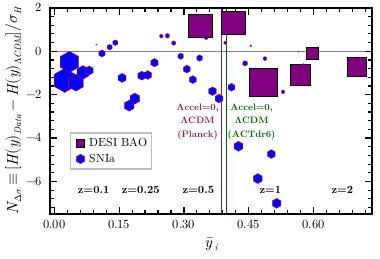}
		\caption{}
		\label{NormedScaledWholePlot}
	\end{subfigure}\hfill
	~ 
	\begin{subfigure}{.475\linewidth}
		\includegraphics[width=\linewidth]{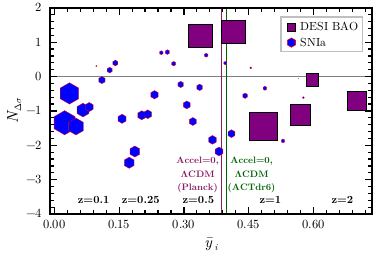}
		\caption{}
		\label{NormedScaledZoomedPlot}
	\end{subfigure}
	\caption{An enhanced version of Fig.~\ref{TwoHofYplots}b, where each of the [$H(y)_{Data} - H(y)_{\Lambda CDM}$] residuals have been divided by their individual $\sigma_{H(y)}$ values, such that the vertical axis now depicts \textit{how many standard deviations} each data point is \textit{above} or \textit{below} this representative $H(y)_{\Lambda CDM}$ model, at the relevant $y$-redshift. (SNIa data shown as blue hexagons, BAO data as purple squares.) To visually emphasize the more reliably-fitted redshift bins (and de-emphasize the doubtful data points), the \textit{size} of each data point $H_{i}$ has been set equal (in the plotting software's graphical scale) to (32/$\sqrt{\sigma _{i}}$). Plot (a) shows a wider view, so that only a single point (at $y = 0.547$) is off the chart; Plot (b) shows the same information, but more zoomed-in, to better illustrate the sharp transition in the BAO data, and the oscillatory pattern apparent in the SNIa data.}
	\label{TwoSCALEDdelHofYplots}
\end{figure*}

Furthermore, these plots are depicted in an enhanced way in Figure~\ref{TwoSCALEDdelHofYplots}, where we have translated the $H(y)$ residuals into a graphic showing the \textit{number of standard deviations} away from the $\Lambda$CDM model for each (SNIa and BAO) data point, by using the vertical axis to plot $N_{\Delta \sigma} \equiv  \{[H(y)_{Data} - H(y)_{\Lambda CDM}]/\sigma _{H(y)} \}$. However, since this effectively removes error bars from the figure, it is no longer obvious which the ``bad" data points are (i.e., the ones with large $\sigma _{H(y)}$ values). To restore some depiction of this data point ``quality" information, we therefore individually scale the plot marker size for each data point $d_{i}$ by the inverse-square-root of its uncertainty value: $Size _{i} \propto 1/\sqrt{\sigma _{i}}$; this graphical scaling choice makes it clear which points should be taken more seriously than others, without shrinking too many of the data points so much as to make them barely visible. Fig's.~\ref{NormedScaledWholePlot} and \ref{NormedScaledZoomedPlot} both show the same information, though the former has a wider range in the vertical axis, in order to contain (almost) all of the points in the chart; while the latter zooms in more, to better depict the interesting physical trends visible in the plot. Vertical lines have also been included to denote the transition redshift -- defined earlier as occurring at the moment of coasting, $\ddot{a} = 0$ -- as estimated by parameters taken from the Planck ($y_{Tr} = 0.3877$) and ACT DR6 ($y_{Tr} = 0.3986$) observations, respectively.

A number of interesting observations can be made based upon Figures~\ref{TwoHofYplots}-\ref{TwoSCALEDdelHofYplots} (and particularly Fig.~\ref{NormedScaledZoomedPlot}). First, focusing upon the DESI DR2 BAO data (purple squares), it is evident (as noted in Sec.~\ref{SecBAOdata}) that there is a sharp jump in $H(y)$ located slightly before the deceleration-to-acceleration transition; this echoes our expectations from Sec.~\ref{SecIntro}, as the lead-up to the onset of acceleration is exactly when the ``dark energy" first grows to become a significant component of the cosmic contents, and its detailed physical nature first becomes important.

Quantitatively,  we noted the transition redshift to be $y_{Tr} \sim 0.39-0.40$ ($z_{Tr} \sim 0.63-0.66$). Consulting Table~\ref{TableHyBAOnumbers}, and going from the BAO data point at $z_{\textrm{eff}} = 0.934$ to the next one at $z_{\textrm{eff}} = 0.706$, we see that the deviation of the data from the $\Lambda$CDM baseline model jumps from $\Delta_{H(y)} = [H(y)_{BAO} - H(y)_{\Lambda CDM}] = -1.8$, up to $+2.3$; this is an abrupt change (in the offset from $\Lambda$CDM) of $+4.1$~km~sec$^{-1}$ Mpc$^{-1}$. More broadly, if we compare the average of the four BAO data points with $z_{\textrm{eff}} > 0.93$ versus the average of the two points with $z_{\textrm{eff}} < 0.71$, we see a \textit{step change} in $\Delta_{H(y)}$ of $\simeq +3.9$ (which increases to a step of $+4.3$~km~sec$^{-1}$ Mpc$^{-1}$ if we neglect the BAO data point with the highest $\sigma_{H(y)}$ at $z_{\textrm{eff}} = 1.484$). Now, whatever compromise is eventually reached between the Hubble Tension estimate from SH0ES of $[H(y)_{SNIa} - H(y)_{CMB}] \simeq 6$, and the estimate from CCHP of $\simeq 3$~km~sec$^{-1}$ Mpc$^{-1}$, it is apparent that this ``step change" seen in the DESI BAO data near the transition redshift $z_{Tr}$ is substantial enough such that it \textit{essentially accounts for the entire Hubble Tension.} Granted, the uncertainties on the BAO data points ($\sigma_{H(y)} \sim 1.3-2.8 ~\&~ 6.5$) are not so small compared to this step change, so that one must be cautious in drawing any conclusions about attributing the Hubble Tension to a misinterpretation by the $\Lambda$CDM paradigm about whatever species or mechanism is actually driving the acceleration. Nevertheless, this is a very intriguing observation, which will hopefully be followed up by a detailed Hubble Tomography breakdown of $H(y)$ versus redshift -- especially around $z_{Tr}$ -- using future BAO (and other) upcoming data sets.

Next, regarding our Linear Cosmography fits to the supernova data, the uncertainties and scatter are even larger, so we must be even more cautious. The problem is most severe at the high-redshift end, particularly (and unfortunately) where the SNIa and BAO data sets overlap. In Fig's.~\ref{TwoHofYplots},\ref{TwoSCALEDdelHofYplots}, one notices an apparent ``bifurcation" of the $H(y)$ results for $y \gtrsim 0.4$, with some Hubble values roughly in line with baseline $\Lambda$CDM, but with other data points being much lower. This is clearly a result of a bifurcation in the data that is also  evident\footnote{By an odd coincidence, looking not unlike the North and South Forks of Long Island, separated by a ``gap" like Shelter Island Sound.} in Fig.~\ref{AllFitsFitFnPlot}, which is not obviously the fault of either the DES5Yr or Union2.1 SNIa data sets, but generally seems present in both. The unfortunate gap in the middle of the high-redshift data leads to a number of physically unlikely Hubble Parameter estimates there (particularly the few negative $H(y)$ values), making any conclusions difficult. But despite this limitation, one can see that the ``best" (smallest $\sigma_{H(y)}$, largest visual point size) data values in Fig.~\ref{NormedScaledWholePlot} do in fact seem to echo the BAO result of a \textit{sharp increase} (relative to $\Lambda$CDM) in $H(y)$ in the lead-up to transition redshift $y_{Tr}$; though in this case, we must take such evidence with a larger grain of salt.

Unfortunately, these figures show that the low-but-sharply-increasing values of $H(y)$ from the supernovae are inconsistent (much lower)\footnote{Despite the common calibration already of SNIa and BAO data here to $H_{0} = 70$~km~sec$^{-1}$ Mpc$^{-1}$, one may think it still possible for the supernovae in Fig's.~\ref{TwoHofYplots},\ref{TwoSCALEDdelHofYplots} to be brought up to the same level as the BAO data with some moderate correction of the ``standardized" SNIa magnitude; but this is not so simple, due to the derived quantities being plotted here. In fact, one would have to reduce the standard SNIa magnitude by 90\% or more to get the SNIa and BAO data points mutually in line at redshifts where these data sets overlap.} than those from the BAO data points (especially obvious in Figs.~\ref{TwoSCALEDdelHofYplots}a,b). A similar inconsistency between the SNIa and BAO data was noted by \citet{HubbTensTomog} -- using a process they termed Hubble \textit{Tension} Tomography -- where they directly binned the distance modulus data $\mu$ (not our Fitting Function $F_{fit} (y)$), and alternatively used a variety of (pre-2025) BAO data sets, and the Pantheon+ supernova compilation. (Our step-change at $y \sim y_{Tr}$ is reminiscent of their ``$\Lambda_{\textrm{s}}$CDM" model.) Their Fig.~1 shows a wide gap everywhere between their binned SNIa points and the BAO data points, as we generally see here. (Using various integrated cosmological models, they claim a larger discrepancy at low redshift, which lessens as redshift increases; yet there does not appear to be \textit{any} redshift region in their figure where the BAO and SNIa points come particularly close to overlapping.) Until a larger, high-quality sample of SNIa data becomes available at higher redshifts, the BAO data remain stronger at elucidating the evolution of the expansion for $y \gtrsim 0.4$; though as redshift decreases from that cosmic transition region, the SNIa data begin to take over as the best source of $H(y)$ information. 

Heading into this post-transition epoch, we find perhaps our most intriguing result: what appears to be a fairly coherent \textit{oscillation in the Hubble Parameter evolution.} This behavior is evident in Figures~\ref{TwoHofYplots}-\ref{TwoSCALEDdelHofYplots}, and looking closely at Fig.~\ref{NormedScaledZoomedPlot}, one sees what \textit{may} be $H(y)$ ``peaks" near the data points at $y = 0.262$, 0.141, and  0.035; and ``troughs" around the points at $y = 0.381$, 0.173, and perhaps 0.05 \& 0.024. Such ``Hubble Oscillations" would be an exciting find, if they can eventually be verified; yet we merely call them ``hints" here, because of the preceding caveats discussed above, and also because these potential oscillations are not very large: only a few $\sigma$'s either way, with an overall maximum oscillatory extent (peak to trough) of only around $\Delta [N_{\Delta \sigma}] \sim 3 \sigma_{H}$ or so. Yet many of the more reliable data points participate in this behavior, where we may be seeing hints of $\sim$2-3 oscillations in the SNIa data for $y \lesssim 0.4$. 

Previous claims of oscillations in the cosmic expansion rate have been made before, such as in \citet{Ringer2015,Ringer2020}, using a variety of SNIa data sets; although those claims have since come under scrutiny \citep{TheHarvardGuys}. More recently, claims of an oscillating dark energy have been made in \citet{WRongOscillations}, though perhaps somewhat confusingly: their redshift space has been carved into over 100 (largely overlapping) bins, potentially exacerbating the ``look-elsewhere effect", and leading to some conflicting claims (for example, per their explanation, parts of their bins covering $z \in [0.36,0.52]$ seem to be accelerating \textit{and} decelerating). No plots have been provided there to elucidate the matter; and their main deceleration region appears to be $z \in [0.4,0.52]$ -- a redshift region where our binned-SNIa $H(y)$ data points seem to indicate an acceleration far sharper than that of $\Lambda$CDM. So it seems fair to say that previous indications of Hubble Parameter oscillations present a muddled picture.

Perhaps most interesting of the previous results are hints of ``wiggles" in the dark energy equation of state $w(z)$ discussed in \citet{WigglingWz}, using the original Pantheon supernova compilation (along with complementary data sets); and based on earlier work by \citet{WangWiggles}, using JLA supernovae \citep{JLASNe} (and other data sets). Intriguingly, \citet{WigglingWz} quote $z \lesssim 0.67$ for where these wiggles appear to set in, which as we noted is when the acceleration era begins. Unfortunately, their hints are also of marginal significance (all at the $\leq 2.1 \sigma$ level, for different Fourier modes, even after imposing priors), but may turn out to be in line with what we are seeing here.

Despite the still-uncertain observational implications to date, a number of theoretical models can be found that have predicted oscillations (about a base $\Lambda$CDM model) in the expansion rate, such as those found in papers cited by the \citet{TheHarvardGuys} publication. Additionally, we note the ``Causal Backreaction" paradigm developed by one of us \citep{BochnerCausBackSmoothInhomog,BochnerCausBackNewConcord,BochnerCausBackRecNonlin} -- one variety of cosmological backreaction, which produces an apparent acceleration from structure formation (without requiring a ``dark energy") -- where one of the predictions was an oscillating acceleration of the Hubble expansion\footnote{For example, a quote from one of our Causal Backreaction papers: ``...the most reasonable scenario in our paradigm may be some kind of stop-and-go accelerating behavior... This could lead to a situation where the universe is always `riding the edge' of a runaway acceleration, but never quite getting there – perhaps in a permanent state of acceleration that oscillates over time but stays relatively close to the level of acceleration that we see today... with these conditions remaining generically true for the foreseeable cosmological future."}. (Some of the simulations from the most updated version of our paradigm \citep{BochnerCausBackRecNonlin} were also able to predict larger values for the current age of the universe that were nearly as high as $\sim14.5$~Gyr, while still fitting the CMB acoustic scale $\theta_{*}$ competitively well -- a feature which could conceivably help with the potential conflict in standard cosmology \citep[e.g.,][]{LCDMStressTest} due to the JWST discovering ``too massive" galaxies, ``too soon" after the Big Bang, given the presumed age of the universe from $\Lambda$CDM.) Furthermore, as our Causal Backreaction paradigm combines the effects of many local inhomogeneities (to produce an averaged effect on larger scales), it evades any limitations due to the so-called no-go ``theorem" for the Hubble Tension \citep[e.g.,][]{NoGoGuideHWY}, which is claimed to constrain late-time deviations from $\Lambda$CDM for homogeneous models.

\begin{figure*}[ht!]
	\begin{subfigure}{.475\linewidth}
		\includegraphics[width=\linewidth]{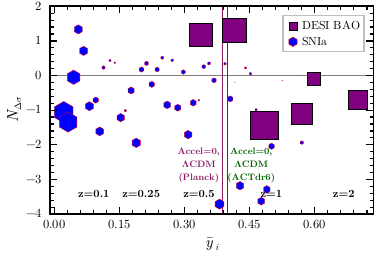}
		\caption{}
		\label{NormedScaled012Plot}
	\end{subfigure}\hfill
	~ 
	\begin{subfigure}{.475\linewidth}
		\includegraphics[width=\linewidth]{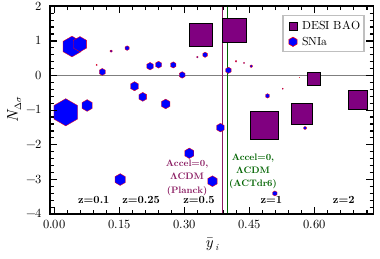}
		\caption{}
		\label{NormedScaled018Plot}
	\end{subfigure}
	\medskip
	\begin{subfigure}{.475\linewidth}
		\includegraphics[width=\linewidth]{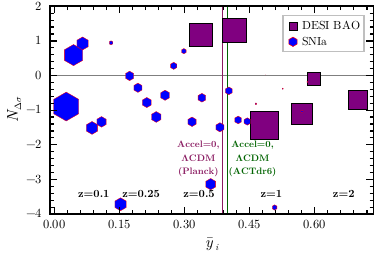}
		\caption{}
		\label{NormedScaled021Plot}
	\end{subfigure}\hfill
	\begin{subfigure}{.475\linewidth}
		\includegraphics[width=\linewidth]{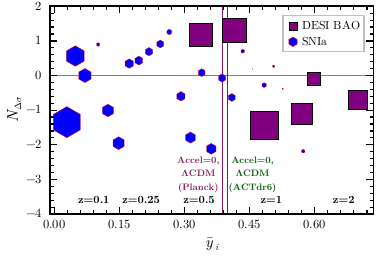}
		\caption{}
		\label{NormedScaled024Plot}
	\end{subfigure}
	\caption{Equivalent plots to that shown in Fig.~\ref{TwoSCALEDdelHofYplots}b, though for the four alternative cases specified in Sec.~\ref{BinningReqs}. Plots (a-d) depict (respectively) the cases with: $\Delta y = \{ 0.012, 0.018, 0.021, 0.024 \}$, where $y_{init} = 0.014$ in every case.}
	\label{FourCasesSCALEDdelHofYplots}
\end{figure*}

As interesting as the phenomenon of an oscillating Hubble Parameter might be, we must ask if this is a robust observation, or merely an artifact of something from this analysis -- such as our chosen binning scheme. We therefore compare by performing a number of runs using the same overall procedure, but with different redshift binnings. As discussed in Sec.~\ref{BinningReqs}, we try alternative binnings with: $\{ y_{init}, \Delta y \} = \{ 0.014, 0.012 \}$, $\{ 0.014, 0.018 \}$, $\{ 0.014, 0.021 \}$, and $\{ 0.014, 0.024 \}$. In Figure~\ref{FourCasesSCALEDdelHofYplots}, we show plots equivalent to that from Fig.~\ref{TwoSCALEDdelHofYplots}b, for each of these four cases. 

Each of these additional cases, in their own way, do appear to show some form of ``oscillations"; though none show them as clearly as our original $\{ y_{init}, \Delta y \} = \{ 0.014, 0.015 \}$ case, and the details of any peaks and troughs are not entirely the same from one binning to another. We therefore cannot conclude that the hints of Hubble Oscillations found earlier are robust against different choices of supernova redshift binnings. It would ultimately be desirable for the volume and quality of SNIa data to be ample to prove (or disprove) such oscillations for \textit{any} reasonable choice of redshift binning scheme; yet, we also note it is possible for a real oscillation to be clearly found -- or almost entirely missed -- based upon whether one has stumbled upon an optimal (or less-than-optimal) sampling frequency. Thus binning-independence is not \textit{necessarily} required as a property of a physically real Hubble Oscillation.

Lastly, we note that an updated version of the Dark Energy Survey supernova compilation has been released fairly recently \citep{DESDovekie}: the DES-Dovekie SNIa data release\footnote{DES-Dovekie SNIa data are currently available online at the former repository for DES5Yr:\\ https://github.com/des-science/DES-SN5YR/tree/main/4$\_$DISTANCES$\_$COVMAT}. This is a substantial update of DES5Yr, involving extensive re-calibrations and associated changes, including substantial alterations to the underlying SNIa sample itself -- with over 100 supernovae from DES5Yr being left out of DES-Dovekie (and vice-versa). The total number of supernovae in their compilation drops from 1829 to 1820, with a similar (in fact, very slightly worse) data gap in the vicinity of $y \sim z \sim 0.1$. The number of duplicated SNIa with Union2.1 also increases from 9 to 10, thus our new $\{$DES-Dovekie + Union2.1$\}$ data set combination now includes a total of $1820 + (580 - 50) = 2350$ distinct supernovae (reduced by 10 SNIa from before).

Most notably, for cosmological modeling with their SNIa (in combination with other data sets), their shift from DES5Yr to Dovekie has \textit{reduced} their tension with the Cosmological Constant (i.e., flat $\Lambda$CDM) from $4.2\sigma$ to $3.2\sigma$ significance. Given this substantial movement, it is thus also interesting to see what effects (if any) may result from this change in the underlying data set for our Hubble Tomography analysis.

\begin{figure*}[ht!]
	\begin{subfigure}{.475\linewidth}
		\includegraphics[width=\linewidth]{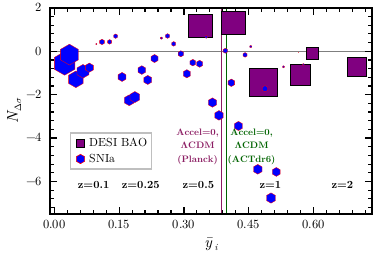}
		\caption{}
		\label{NormedScaledWholeDovekiePlot}
	\end{subfigure}\hfill
	~ 
	\begin{subfigure}{.475\linewidth}
		\includegraphics[width=\linewidth]{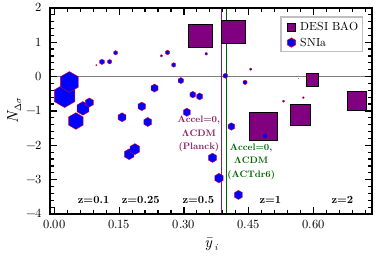}
		\caption{}
		\label{NormedScaledZoomedDovekiePlot}
	\end{subfigure}
	\caption{Equivalent plots to those shown in Fig.~\ref{TwoSCALEDdelHofYplots}, for the same binning scheme -- $\{ y_{init}, \Delta y \} = \{ 0.014, 0.015 \}$ -- but now done using the updated SNIa compilation, DES-Dovekie, in combination with Union2.1.}
	\label{TwoScaledDovekieplots}
\end{figure*}

While we do not repeat our entire Linear Cosmography study from DES5Yr, we have performed a re-analysis of the $\{ y_{init}, \Delta y \} = \{ 0.014, 0.015 \}$ case for Dovekie. The equivalent of the plots in Fig.~\ref{TwoSCALEDdelHofYplots}, redone with the updated supernova compilation, are shown in Figure~\ref{TwoScaledDovekieplots}.

Comparing the two figures, we see that the overall behaviors of the data are roughly the same here for Dovekie as they were for DES5Yr. The apparent $H(y)$ oscillation during the cosmic acceleration period is still present, if perhaps a bit more ragged for the Dovekie data. A somewhat bigger difference occurs around the cosmic transition redshift, where a few significant (moderate $\sigma_{H}$) points around $y \lesssim y_{Tr}$ have gotten ``worse" for Dovekie -- i.e., $H(y)$ getting lower, increasing the discrepancy with DESI BAO (and with baseline $\Lambda$CDM); but at least one similar point around $y \gtrsim y_{Tr}$ has moved up closer towards the DESI level, so it is almost a wash. The overall issue at higher redshifts is still the unreliability of results around $y \sim y_{Tr}$ and above, due to the sparseness and the bifurcation of the data there.

On the whole, our Hubble Tomography method does not show a great difference when exchanging the DES-5Yr SNIa for DES-Dovekie data, other than for a few isolated instances in a few redshift bins. In contrast to \citet{DESDovekie}, where the update is enough to change their evidence for evolving Dark Energy from ``strong" to ``weak", our tomographic approach is more focused (perhaps even myopic), keeping the measurement of the Hubble evolution within each particular redshift region isolated from whatever effects are going on at other redshifts. Though the measurements from our method are not yet statistically strong enough for us to quote quantitative parameter error ellipses, all of the qualitative results discussed in this section for DES5Yr seem to be carried over to the new DES-Dovekie data update, with similar reliability.

\section{Summary and Conclusions \label{SecConclude}} 

In this paper, we have presented an alternative method for measuring the cosmic evolution: Hubble Tomography via Linear Cosmography. Unlike the usual approach of fitting specific theoretical models to combinations of data sets across the entire redshift range of the data, our method uses Type Ia Supernova data to explicitly measure the Hubble Parameter $H(y)$ ($\equiv H(z)$) in piecewise fashion, binned in redshift. Thus we can focus on the detailed temporal behavior of the cosmic expansion, and pinpoint any specific instances or types of deviations from standard models like flat $\Lambda$CDM. This method may also be particularly well suited for \textit{directly} locating and characterizing the source of the controversial Hubble Tension, if it turns out to be a physically real phenomenon.

We began by stating our expectation that any discrepancies versus Cosmological Constant models would most likely first show up in cosmic history around (or slightly before) the transition from deceleration to acceleration, occurring at transition redshift $z_{Tr} \sim 0.63-0.66$ (i.e., $y_{Tr} \equiv [z_{Tr}/(1+z_{Tr})] \sim 0.38-0.40$) -- representing the time when the ``dark energy" grew to become a significant component of the universe, such that its detailed physical nature became important for the very first time. We then noted that while very early (i.e., Cosmic Microwave Background) data is quite consistent (as expected) with $\Lambda$ as the dark energy, it is already apparent in Baryon Acoustic Oscillation data that there is a \textit{sharp jump} in the Hubble Parameter right around this expected redshift range, with a change $\Delta H$ that is nearly equal to the \textit{entire} Hubble Tension (assuming some compromise between its currently estimated values).

Moving on to our linear cosmography analysis of SNIa data, we noted that the recent supernova compilations, being composed of several distinct data sets fused together, consequently have a ``gap" (sparseness) in the data, with the deepest lack of data located around $y \sim z \sim 0.1$ (and broadly covering the $y \sim 0.08-0.16$, $z \sim 0.09-0.19$ redshift range). Having chosen the DES 5-Year Data Release as our main SNIa data set, we therefore needed to supplement this with the Union2.1 supernova compilation, helping to partially fill in that data gap, while also incidentally extending the full range of SNIa data at both the low-redshift and high-redshift ends.

Conducting our Hubble Tomography analysis, we found that there does appear to be a jump (relative to $\Lambda$CDM) in the Hubble Parameter around (\textit{and} before) transition redshift $y_{Tr}$ for the SNIa data also, similar to that seen in BAO data; though unfortunately, the supernova data produces significantly lower values of $H(y)$ there, and is thus not consistent with the DESI BAO data at high redshifts. We attribute this to a ragged sample set at the high-redshift end, where the SNIa plots show literal holes and bifurcations in the data as it trails off. 

But after the epoch of the cosmic transition from deceleration to acceleration, we get our most interesting result: hints of what appears to be an oscillation in the Hubble Parameter, perhaps $\sim$2-3 oscillations in $H(y)$ from $y_{Tr} \sim 0.4$ down to the low-redshift end of the data. This is clearly a tentative result, subject to the list of caveats discussed previously -- most importantly, being an oscillation of an amplitude that only extends $\sim 3 \sigma_{H}$ from peak to trough, and one that does not currently retain the exact same structure for different redshift binnings. However, the result remains intriguing nonetheless, and the abundance of new Type Ia supernovae expected from upcoming surveys should provide further testing of this tentative observation. But regardless of the specific findings of this analysis, the method of Hubble Tomography introduced here will hopefully find a continuing use -- and would become much more definitive, assuming that the uncertainties from photometrically-measured redshifts can be controlled well enough -- when those more comprehensive SNIa data sets arrive.

Finally, we note again that an oscillation in the Hubble expansion evolution, though generally a nonstandard idea for cosmological modeling, has in fact been predicted by the theories of a number of researchers (including one of us). Most importantly, such a finding (if it bears out), would mark a fundamental change in our cosmological outlook. After roughly a quarter-century of the (arguably metaphysical) approach of circumscribing the expected density of Cosmological Constant dark energy using Anthropic arguments in multiverse theories, there is nothing more typical of physics than returning to the basic idea of an oscillating dark energy: a battle of cosmic force and restoring counter-force, representing the most commonly observed type of behavior in our collective scientific experience.

\vspace{1cm}

\appendix

\section{Two Sides of the Same Coin: The Hubble Tension, and Deviations from $\Lambda$CDM \label{BothSameProb}}

In Section~\ref{SecIntro} it was noted that the possible ``Concordance problems" of the Hubble Tension, and deviations from standard $\Lambda$CDM, are not actually two independent problems (as they may commonly be considered to be), but are actually strongly correlated issues.

\begin{figure}[h!]
	\centering
	\epsscale{0.50}
	\plotone{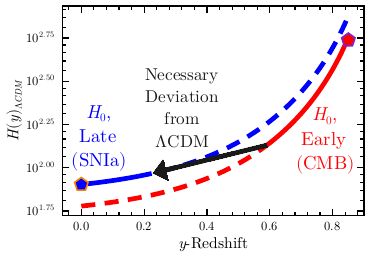}
	\caption{Schematic demonstration of how a Hubble Tension implies a deviation from the standard cosmological model, using a pair of $\Lambda$CDM toy models with $\Omega_{M} = (1 - \Omega_{\Lambda}) = 0.3$: $H^{Early}_{0} \equiv 60$~km~sec$^{-1}$ Mpc$^{-1}$ \textit{\{bottom line\}}, and, $H^{Late}_{0} \equiv 80$~km~sec$^{-1}$ Mpc$^{-1}$ \textit{\{top line\}}. The filled pentagons represent the starting and ending points for the cosmic evolution in this figure, which always follows the solid line segments by ``switching tracks", an action requiring a deviation from pure-$\Lambda$CDM behavior. \label{HubbTensionDevPlot}}
\end{figure}

On the one hand, note that measurements like CMB observations (or any other high-redshift proxy) do not actually measure the Hubble Constant $H_{0}$, but merely measure the integrated distance to the object of study, such that the temporal details of the expansion rate are ``smoothed" over a long period during which $H(z)$ has significantly evolved. This distance is \textit{proportional} to $H_{0}^{-1}$, but some model for the cosmic scale factor $a(t)$ must be \textit{assumed} to turn this into a quantitative constraint on $H_{0}$. In the context of Concordance parameter estimations, this model is always $\Lambda$CDM\footnote{To quote from \citet{PlanckH0}: ``...the values given [of $H_{0}$ and $\Omega_{M}$] assume the base-$\Lambda$CDM cosmology...These estimates are highly model dependent and this needs to be borne in mind when comparing with other measurements, for example the direct measurements of $H_{0}$..."},\footnote{In a similar vein, claims of fundamental (``no-go") limitations to late-time (post-recombination) solutions for the Hubble Tension -- see, e.g., \citet{SunnyVagNoLate}, and references therein -- are also assumption-dependent. (Such as assuming the ongoing validity of the FRW homogeneity assumption, and that terms like ``$\Omega_{M}$" mean the same thing at late times, as they did during the CMB epoch.)}. In short, if the value of $H_{0}$ actually measured via low-redshift standard candle/ruler/siren measurements \textit{does not match} the value(s) inferred (using $\Lambda$CDM) from indirect high-redshift measurements -- and if all observations (including CMB estimates of $\Lambda$CDM parameters besides $H_{0}$) are assumed to be correct -- then basic continuity of the $a(t)$ function \textit{requires} there to be a deviation from $\Lambda$CDM at some point. A schematic visualization of this is shown in Figure~\ref{HubbTensionDevPlot}.

In other words, a known value of $H_{0}$, plus a specific $\Lambda$CDM model, \textit{predicts} the distance to the CMB last scattering surface -- or to any other high-redshift object being studied -- and a ``mismatch in $H_{0}$" is another way of saying that the $\Lambda$CDM-predicted \textit{distance} is wrong. (Theoretical fiddling with the non-$\Lambda$ cosmic contents -- e.g., non-standard dark matter, relativistic species, etc.\ -- can provide some leeway in the matter, though to a limited degree.) Since the distance to the CMB is observationally measured to relatively high precision via the acoustic scale $\theta_{*}$ from the angular size of the CMB peaks, this implies a necessary deviation from $\Lambda$CDM to match the CMB distance measurement.

Conversely, a deviation from $\Lambda$CDM evolution pretty strongly implies that a mismatch of $H_{0}$ estimates (i.e., a Hubble Tension) is likely to be observed in high-$z$ versus low-$z$ measurements, since it deflects the cosmic evolution to a ``different track", so to speak, as depicted in Figure~\ref{HubbTensionDevPlot}. (It is of course \textit{possible} for multiple or complicated deviations to balance each other and integrate out in a way that leaves the final value of $H_{0}$ unchanged from the $\Lambda$CDM expectation -- similar to how ``Mirage of w=-1" models \citep{LinderMirage,DESIMirage} cancel out in a way to hide deviations in the dark energy equation of state $w(z)$ evolution -- but it would seem to be a rather coincidental outcome for a Hubble Tension to be ``hidden" in this way.)

\vspace{1cm}

\section{Replacing $z$-Redshift with $y$-Redshift in Theoretical Cosmology \label{AppRedshift}}

As noted in Section~\ref{yVSzRedshift} above, we advocate for the abandonment of the cosmological use of traditional $z$-redshift, in favor of the mathematically more justifiable $y$-redshift. Here we give the background for our arguments.

Historically, from an observational point of view, one uses $z = [a_{0} - a(t)]/a(t) = (\Delta \lambda / \lambda_{em})$ -- i.e., the redshift-induced change in wavelength divided by the \textit{emitted} wavelength -- which is determined by the astrophysics of the source. Thus $z$ clearly indicates the region of the spectrum in which we expect to find a particular cosmologically-distant signal, hence it was always the crucial parameter in terms of telescope technology selection. Also, in the early works measuring the Hubble Constant -- when only fairly low-redshift data was available -- the approach was to plot velocity versus distance ($v = H_{0} \times D$), and for low redshifts we of course have $v \simeq c z$, so that $z$-redshift served as the effective ``$y$-axis" of those early Hubble plots (before modern cosmology promoted redshift to the $x$-axis, as the ``independent" variable).

However, with data now available from much greater distances, modeling this data from the present time back to the Big Bang (or significantly closer to it), we have redshift range $z \in [0,\infty)$, which causes several problems. First, $z$-redshift dramatically ``stretches out" early universe epochs relative to more recent times, making it a highly nonlinear and non-representative parameter for proportionally representing cosmic time $t$ (or for paralleling the development of structure formation). More seriously, as has been quite effectively pointed out \citep[see][and their series of papers]{CattVissYred}, cosmological fits using Taylor-like expansions will fail to sensibly converge when applied to high-redshift data with $z \gtrsim 1$ -- a threshold greatly exceeded already by current SNIa and BAO data sets.

The alternative $y$-redshift, however, is equal to the redshift-induced change in wavelength divided by the \textit{observed} wavelength, satisfying $y = (\Delta \lambda / \lambda_{obs}) = 1 - [a(t)/a_{0}]$. Thus it has range $y \in [0,1]$ going back to the Big Bang (clearly more valid for series convergence); and it is better suited for parameterizing the evolution of the universe -- obviously so in terms of structure formation, particularly in the linear regime, where $\delta \rho / \rho \propto a(t)$ \citep{KolbTurner}. Also, while $y$-redshift is not \textit{precisely} linear in cosmic time $t$, it is far closer to linear in $t$ than $z$-redshift happens to be, for realistic cosmological models.

For many recent cosmological studies, this is a moot point, since series expansions are often done now using the CPL parameterization \citep{ChevPolCPL,LindCPL} for the dark energy -- i.e., in ``$w_{0}-w_{a}$ space" -- where the expansion variable is \textit{already} defined as $[z/(1+z)] \equiv y$. However, results are often plotted and discussed as functions of $z$, which is unnecessarily misleading for the task of providing barometers of cosmic history. Even worse, one sometimes still finds \textit{cosmographic} studies being done using expansions in $z$-redshift \citep[e.g.,][]{zRedshiftCosmog}, which -- given the very-high-redshift ($z > 2$) data being included in the fits -- is mathematically poor practice. 

Another serious problem is that all cosmographic series expansions in $z$ have the undesirable asymptotic behavior of blowing up towards positive or negative infinity -- the direction depending upon the sign of the highest-order term -- as $z \rightarrow \infty$, leading to unphysical trends at the high-redshift end of the data sample. One result of this, is that fitting with curves using high powers of $z$-redshift forces the so-called ``best-fit" curves to be anchored most strongly by the highest-redshift data, which often tends to be the most unreliably measured data (in a SNIa compilation, for example).

This unnatural constraint at the high-redshift end forces the best-fit cosmographic curves to ``buckle" oddly, disturbing the resulting details of the reconstructed cosmic evolution, and likely also the parameter value estimates ($H_{0}$, $q_{0}$, $j_{0}$, etc.). For example, for the cosmographic fit shown in Figure 2 of \citet{zRedshiftCosmog} -- where the two highest redshift data points are from DESI-BAO -- one can see that the cosmographic fit is forced to go \textit{almost exactly through} those two central data points themselves, not merely through their $\sim \pm 1 \sigma$ ranges, which visibly bends the entire curve out of shape. This problem may in fact explain the caveat they make in their statement about their cosmographic best-fit, that: ``...the expansion history differs substantially from that of the best fit Planck model... it remains important to consider that there might be an unexposed systematic
error."

A cautionary note about the potential change of redshift variables in cosmography was sounded by \citet{CosmogNoGood}, where they showed how using $y$-redshift as the expansion variable could provide less accurate (and biased) estimates of the cosmographic parameters, in comparison to $z$-redshift cosmography. This is not entirely unexpected: because $y$ is always a smaller expansion variable than $z$, its expansion coefficients (and their errors) will be bigger \citep{WigglingWz}; and the bias may be due to the fact that $y$-redshift (particularly in simple expansions) is known to be a ``stiff" parameter, which can act like a prior by restricting the ``wandering" of the equation of state -- see \citet{StiffyRed} (especially their Fig. 11 and the related discussion), and \citet{BochnerLamDev}.

However, the study in \citet{CosmogNoGood} was based on data with a redshift distribution modeled on that of Union2.1, for which only 5\% of the data has $z > 1$, and their Fig. 1 plot demonstrating the relative success of $z$-cosmography only goes up to $z = 1$ (beyond which the expansion would make no sense, regardless of how small the estimated parameter uncertainties might seem to be -- precision is not accuracy). In contrast, the DESI DR2 BAO data set has points that go all the way up to $z_{eff} = 2.33$, and this data is used both here, and in \citet{zRedshiftCosmog}, with those few highest-$z$ points dominating their fit (as is bound to happen in $z$-based expansions).

Yet, the conclusions reached by \citet{CosmogNoGood} are well-taken, and perhaps the productive use of cosmography is a difficult challenge with any expansion variable. But this is precisely why we focus here upon \textit{linear} cosmography exclusively, applied to bin sizes small enough such that the Hubble Parameter can be modeled with straight line segments by any cosmology that is not varying extraordinarily rapidly. And even in this context, $z$-redshift fails as a fitting variable. With $y = 1 - a$ (letting $a_{0} \equiv 1$), we have $|\Delta y|$ = $\Delta a$, which is a fairly stable binning variable; but with $z = (1/a) - 1$, and thus $|\Delta z| = (\Delta a / a^{2})$, we see that as one goes back in time and $a(t)$ becomes small, this relationship hugely stretches out the $\Delta z$-widths of bins representing very small slices of cosmic history.

Therefore, we conduct our analysis in terms of $y$-redshift everywhere; yet, most results here will also be re-interpreted in terms of $z$-redshift, just to be more understandable for the common cosmological discussion. However, we believe that the best practice would be to overcome the inertia of using $z$-redshift, and replace it entirely with $y$-redshift -- or even simply with $(1 - y) = (1 + z)^{-1} = ( \lambda_{em} / \lambda_{obs} ) = [a(t)/a_{0}]$ -- for all serious discussions in theoretical cosmology.

\vspace{0.5cm}

\begin{acknowledgments}
	Note that we used the \textit{Julia} programming language \citep{JuliaRef} for the numerical fitting work and plots in this paper; much help with coding and figure formatting was received from the Julia Discussion Board (https://discourse.julialang.org). We are grateful to Willem Elbers, David Rubin, and Brodie Popovic for useful feedback about their data sets. We also thank Sunny Vagnozzi and Eoin \'{O} Colg\'{a}in for their helpful comments in response to the first posted version of this paper.
\end{acknowledgments}


\vspace{1cm}



\begin{thebibliography}{}

\bibitem[Abbott et al.(2024)]{DES5YrSNIa}
Abbott, T. M. C., et al. (DES Collaboration)
2024, \apjl~973(1), L14. 
https://arXiv.org/abs/2401.02929

\bibitem[Abdul-Karim et al.(2025)]{DESIBAO}
Abdul-Karim, M., et al. (DESI Collaboration)
2025, Phys. Rev. D~112, 083515.
https://arXiv.org/abs/2503.14738

\bibitem[Aghanim et al.(2020)]{PlanckH0}
Aghanim, N., et al. (Planck Collaboration)
2020, Astron. \& Astrophys.~641, A6.
https://arXiv.org/abs/1807.06209

\bibitem[Akarsu et al.(2024)]{LocH0Tensions}
Akarsu, \"{O}., Colg\'{a}in, E. \'{O}, Sen, A. A. \& Sheikh-Jabbari, M. M.
2024, Universe~10(8), 305
https://arXiv.org/abs/2402.04767

\bibitem[Bansal \& Huterer(2025)]{DragonH}
Bansal, P. \& Huterer, D.
2025, Phys. Rev. D~112(2), 023528.
https://arXiv.org/abs/2502.07185

\bibitem[Betoule et al.(2014)]{JLASNe}
Betoule, M., et al. (SNLS-SDSS ``JLA" analysis)
2014, Astron. \& Astrophys.~568, A22.
https://arXiv.org/abs/1401.4064

\bibitem[Bezanson et al.(2017)]{JuliaRef}
Bezanson, J., Edelman, A., Karpinski, S. \& Shah, V.
2017, SIAM review~59(1), 65.

\bibitem[Bochner(2011a)]{BochnerCausBackSmoothInhomog}
Bochner, B. 
2011a, https://arXiv.org/abs/1109.4686

\bibitem[Bochner(2011b)]{BochnerCausBackNewConcord}
Bochner, B. 
2011b, https://arXiv.org/abs/1109.5155

\bibitem[Bochner(2013)]{BochnerCausBackRecNonlin}
Bochner, B. 
2013, Int. J. Mod. Phys. D~22(13), 1330026.
https://arXiv.org/abs/1206.5056

\bibitem[Bochner(2025)]{BochnerGR24}
Bochner, B. 2025, “Probing the Temporal Fine Structure 
of Hubble Parameter Evolution with Linear Cosmography”, 
Accepted for Publication in the \textit{Proceedings of the 
24th International Conference on General Relativity and 
Gravitation} (Glasgow, Scotland, July 2025).

\bibitem[Bochner, Pappas \& Dong(2015)]{BochnerLamDev}
Bochner, B., Pappas, D., \& Dong, M.
2015, \apj~814(1), 7. https://arXiv.org/abs/1308.6050

\bibitem[Bousis \& Perivolaropoulos(2024)]{HubbTensTomog}
Bousis, D. \& Perivolaropoulos, L.
2024, Phys. Rev. D~110(10), 103546.
https://arXiv.org/abs/2405.07039

\bibitem[Boylan-Kolchin(2023)]{LCDMStressTest}
Boylan-Kolchin, M.
2023, Nat. Astron.~7, 731.
https://arXiv.org/abs/2208.01611

\bibitem[Brownsberger et al.(2020)]{TheHarvardGuys}
Brownsberger, S. R., Stubbs, C. W. \& Scolnic, D. M.
2020, \mnras~498(4), 5512.
https://arXiv.org/pdf/2005.08029

\bibitem[Busti et al.(2015)]{CosmogNoGood}
Busti, V. C., del la Cruz-Dombriz, \'{A}., Dunsby, P. K. S. \& Saez-Gomez, D.
2015, Phys. Rev. D~92, 123512.
https://arXiv.org/abs/1505.05503

\bibitem[Calabrese et al.(2025)]{ACT6noEDE}
Calabrese, E., et al. (Atacama Cosmology Telescope)
2025, in press (JCAP).
https://arXiv.org/abs/2503.14454

\bibitem[Camilleri et al.(2025)]{zRedshiftCosmog}
Camilleri, R., et al. (DES Collaboration)
2025, \mnras~537(2), 1818.
https://arXiv.org/abs/2406.05049

\bibitem[Catto\"{e}n \& Visser(2007)]{CattVissYred}
Catto\"{e}n, C. \& Visser, M.
2007, Class. Quantum Grav.~24, 5985. 
https://arXiv.org/abs/0710.1887

\bibitem[Chevallier \& Polarski(2001)]{ChevPolCPL}
Chevallier, M. \& Polarski, D.
2001, Int. J. Mod. Phys. D~10, 213.
https://arXiv.org/abs/gr-qc/0009008

\bibitem[Colg\'{a}in et al.(2021)]{WigglingWz}
Colg\'{a}in, E. \'{O}, Sheikh-Jabbari, M. M. \& Yin, L.
2021, Phys. Rev. D~104(2), 023510.
https://arXiv.org/abs/2104.01930

\bibitem[de Bernardis et al.(2000)]{BoomFlat}
de Bernardis, P., et al. (BOOMERanG Collaboration)
2000, Nature~404, 955. https://arXiv.org/abs/astro-ph/0004404

\bibitem[Efstathiou(2025)]{EfStathPanthOff}
Efstathiou, G. 
2025, \mnras~538(2), 875.
https://arXiv.org/abs/2408.07175

\bibitem[Freedman et al.(2025)]{FreedmanH0}
Freedman, W. L., et al.  (Chicago-Carnegie Hubble Program)
2025, \apj~985 (No. 2), 203.
https://arXiv.org/abs/2408.06153

\bibitem[Freedman \& Madore(2023)]{FreedmanCephSys}
Freedman, W. L. \& Madore, B. F.
2023, JCAP~11, 050. https://arXiv.org/abs/2309.05618

\bibitem[Hawking \& Ellis(1973)]{HawkingEllis} 
Hawking, S. W., \& Ellis, G. F. R.
1973, The large scale structure of space-time 
(Cambridge University Press: Cambridge, UK)

\bibitem[Huang et al.(2025)]{NoGoGuideHWY}
Huang, L., Wang, S.-J. \& Yu, W.-W.
2025, Sci. China Phys. Mech. Astron.~68, 220413.
https://arXiv.org/abs/2401.14170

\bibitem[Ivesi\'{c} et al.(2025)]{SNeFromLSST}
Ivesi\'{c}, \v{Z}., et al. (The Large Synoptic Survey Telescope)
2019, \apj~873 (No. 2), 111. https://arXiv.org/pdf/0805.2366

\bibitem[Kolb \& Turner(1990)]{KolbTurner} 
Kolb, E. W., \& Turner, M. S. 1990, The Early Universe 
(Addison-Wesley: Redwood City, CA)

\bibitem[L'Huillier et al.(2019)]{SmoothShaf}
L'Huillier, B., Shafieloo, A., Linder, E. V. \& Kim, A. G.
2019, \mnras~485(2), 2783.
https://arXiv.org/abs/1812.03623

\bibitem[Linder(2003)]{LindCPL}
Linder, E. V. 2003, Phys. Rev. Lett.~90, 091301
https://arXiv.org/abs/astro-ph/0208512

\bibitem[Linder(2007)]{LinderMirage}
Linder, E. V. 2007, https://arXiv.org/abs/0708.0024

\bibitem[Lodha et al.(2025)]{DESIMirage}
Lodha, K., et al. (DESI Collaboration)
2025, Phys. Rev. D~112, 083511. 
https://arXiv.org/abs/2503.14743

\bibitem[Louis et al.(2025)]{ACTDR6}
Louis, T., et al. (Atacama Cosmology Telescope)
2025, in press (JCAP).
https://arXiv.org/abs/2503.14452

\bibitem[Moskowitz et al.(2024)]{PhotozFromLSST}
Moskowitz, I., et al. (The LSST Dark Energy Science Collaboration)
2024, \apjl~967, L6. 
https://arXiv.org/abs/2402.15551

\bibitem[Ostriker \& Steinhardt(1995)]{OstrikerConcord}
Ostriker, J. P. \& Steinhardt, P. J. 
1995, https://arXiv.org/abs/astro-ph/9505066

\bibitem[Pedrotti et al.(2026)]{SunnyVagNoLate}
Pedrotti, D., Escamilla, L. A., Marra, V., Perivolaropoulos, L. \& Vagnozzi, S.
2026, Phys. Rev. D~113, 043507.
https://arXiv.org/abs/2510.01974

\bibitem[Perivolaropoulos \& Skara(2022)]{ChallengesLCDM}
Perivolaropoulos, L. \& Skara, F.
2022, New Astron. Rev.~95, 101659.
https://arXiv.org/abs/2105.05208

\bibitem[Perlmutter et al.(1999)]{PerlAccel99}
Perlmutter, S., et al. (The Supernova Cosmology Project)
1999, \apj~517, 565.
https://arXiv.org/abs/astro-ph/9812133

\bibitem[Popovic et al.(2025)]{DESDovekie}
Popovic, B., et al. (DES Collaboration)
https://arXiv.org/abs/2511.07517

\bibitem[Riess et al.(1998)]{RiessAccel98}
Riess, A. G., et al. (The High-Z Supernova Search Team)
1998, Astron. J.~116, 1009. https://arXiv.org/abs/astro-ph/9805201

\bibitem[Riess et al.(2007)]{StiffyRed}
Riess, A. G., et al.
2007, \apj~659, 98. 
https://arXiv.org/abs/astro-ph/0611572

\bibitem[Riess et al.(2022)]{RiessH0}
Riess, A. G., et al. (The SH0ES Team)
2022, \apjl~934, L7. 
https://arXiv.org/abs/2112.04510

\bibitem[Ringermacher \& Mead(2015)]{Ringer2015}
Ringermacher, H. I. \& Mead, L. R. 
2015, \aj~149(4), 137.
https://arXiv.org/pdf/1502.06140

\bibitem[Ringermacher \& Mead(2020)]{Ringer2020}
Ringermacher, H. I. \& Mead, L. R.
2020, \mnras~494(2), 2158.
https://arXiv.org/pdf/1901.10311

\bibitem[Rubin et al.(2025)]{Union3Rubin}
Rubin, D., et al. (UNITY1.5 Framework)
2025, AAS, 245, 250.02.
https://arXiv.org/abs/2311.12098

\bibitem[Scolnic et al.(2018)]{PantheonOrigData}
Scolnic, D., et al. (Pantheon Analysis)
2018, \apj~859, 101. 
https://arXiv.org/abs/1710.00845

\bibitem[Scolnic et al.(2022)]{PantheonPlusScol}
Scolnic, D., et al. (Pantheon+ Analysis)
2022, \apj~938, 113. 
https://arXiv.org/abs/2112.03863

\bibitem[Shafieloo et al.(2006)]{ShafSmooth}
Shafieloo, A., Alam, U., Sahni, V. \& Starobinsky, A. A.
2006, \mnras~366(3), 1081.
https://arXiv.org/abs/astro-ph/0505329

\bibitem[Suzuki et al.(2012)]{SCPUnion21}
Suzuki, N., et al. (The Supernova Cosmology Project)
2012, \apj~746, 85. https://arXiv.org/abs/1105.3470

\bibitem[Turner(2022)]{TurnerPrecCos}
Turner, M. S.
2022, Annu. Rev. Nucl. Part. Sci.~72, 1.
https://arXiv.org/abs/2201.04741

\bibitem[Valentino et al.(2021)]{HubbTension}
Valentino, E. D., Mena, O., Pan, S., Visinelli, L. \& Yang, W., 
Melchiorri, A., Mota, D. F., Riess, A. G. and Silk, J. 
2021, Class. Quantum Grav.~38, 153001. 
https://arXiv.org/abs/2103.01183

\bibitem[Valentino et al.(2025)]{CosmoVerse}
Valentino, E. D., et al. (The CosmoVerse Network)
2025, Phys. Dark Univ. 49, 101965.
https://arXiv.org/abs/2504.01669

\bibitem[Vincenzi et al.(2024)]{DESnoSystematics}
Vincenzi, M., et al. (DES Collaboration)
2024, \apj~975(1), 86.
https://arXiv.org/abs/2401.02945

\bibitem[Wang et al.(2018)]{WangWiggles}
Wang, Y., Pogosian, L., Zhao, G. B. \& Zucca, A.
2018, \apjl~869(1), L8. 
https://arXiv.org/pdf/1807.03772

\bibitem[Yang(2026)]{WRongOscillations}
Yang, R.-J.
2026, Res. Astron. Astrophys.~26, 015003.
https://arXiv.org/abs/2505.00058

\bibitem[Zhou et al.(2025)]{PreventsLateH0SOlv}
Zhou, Z., Miao, Z., Sheng, B., Ai, C. \& Zhang, H.
2025, Phys. Rev. D~112, 103502. 
https://arXiv.org/abs/2506.23556

\end{thebibliography}
\end{document}